\UseRawInputEncoding
\documentclass[twocolumn]{aastex62}
\usepackage{graphicx}
\usepackage{amssymb}
\usepackage{epsfig}
\usepackage{bm}
\usepackage{color}
\usepackage{amsmath}
\usepackage{natbib}
\usepackage{cancel}
\usepackage{soul}
\usepackage{mathrsfs}
\usepackage{ulem}

\newcommand{\rmf}[1]{{_{\rm #1}}}

\received{2020 November 23}
\revised{2021 January 29}
\accepted{2021 February 10}
\submitjournal{ApJ}
\shortauthors{Ohmura {\it et al.}}
\begin{document}
\title{Continuous Jets and Backflow Models for the Formation of W50/SS433 in Magnetohydrodynamics Simulations}
\correspondingauthor{T. Ohmura} \email{ohmura@phys.kyushu-u.ac.jp}
\author[0000-0002-0040-8968]{T. Ohmura}
\affiliation{Department of Physics, Faculty of Sciences, Kyushu University, 744 Motooka, Nishi-ku, Fukuoka 819-0395, Japan}
\author{K. Ono}
\affiliation{4-10-23, Hayamiya, Nerima-ku, Tokyo, 179-0085, Japan}
\author[0000-0002-4037-1346]{H. Sakemi}
\affiliation{Department of Physics, Faculty of Sciences, Kyushu University, 744 Motooka, Nishi-ku, Fukuoka 819-0395, Japan}
\author[0000-0001-8078-9221]{Y. Tashima}
\affiliation{Department of Physics, Faculty of Sciences, Kyushu University, 744 Motooka, Nishi-ku, Fukuoka 819-0395, Japan}
\author[0000-0002-7259-1706]{R. Omae}
\affiliation{Department of Physics, Faculty of Sciences, Kyushu University, 744 Motooka, Nishi-ku, Fukuoka 819-0395, Japan}
\author[0000-0001-6353-7639]{M. Machida}
\affiliation{Division of Science, National Astronomical Observatory of Japan, 2-21-1 Osawa, Mitaka, Tokyo 181-8588, Japan}
\begin{abstract}
The formation mechanism of the W50/SS433 complex has long been a mystery.
We propose a new scenario in which the SS433 jets themselves form the W50/SS433 system.
We carry out magnetohydrodynamics simulations of two-side jet propagation using the public code CANS+.
As found in previous jet studies, when the propagating jet is lighter than the surrounding medium, the shocked plasma flows back from the jet tip to the core.
We find that the morphology of light jets is spheroidal at early times, and afterward, the shell and wings are developed by the broadening spherical cocoon.
The morphology strongly depends on the density ratio of the injected jet to the surrounding medium.
Meanwhile, the ratio of the lengths of the two-side jets depends only on the density profile of the surrounding medium.
We also find that most of the jet kinetic energy is dissipated at the oblique shock formed by the interaction between the backflow and beam flow, rather than at the jet terminal shock.
The position of the oblique shock is spatially consistent with the X-ray and TeV gamma-ray hotspots of W50.
\end{abstract}
\section{Introduction} \label{sec:intro}
Astrophysical jets emitted from gravitational objects such as black holes, neutron stars, and protostars are universal phenomena observed in various layers of the Milky Way.
The jets interact with the surrounding interstellar medium (ISM), and the kinetic energy of the jets is therefore transformed into ISM thermal and turbulent energies.
Additionally, some of the kinetic energy of jets is given to energetic, non-thermal particles.
Microquasar jets are therefore a candidate for cosmic-ray acceleration sites \citep[e.g.,][]{2002A&A...390..751H,2009A&A...497..325B}.
In particular, \cite{2020MNRAS.493.3212C} argued that the cosmic-ray component generated in microquasar jets makes a non-negligible contribution to observed cosmic-ray spectra.
However, physical quantities and characteristics of galactic jets are currently not well understood because the angular size of many galactic jets is small compared with the spatial resolution of observations.

The radio nebula W50 is located near the galactic plane, and it is a rare galactic object for which high-quality radio observations have been made \citep{dubner1998,2011A&A...529A.159G,farnes2017,broderick2018}.
The radio morphology of W50 has two characteristics, namely a circular shell with a diameter of 58 arcmin and lateral extensions, which are called wings, in east–west directions.
The X-ray binary SS433 launches sub-relativistic precessing jets at the center of the shell structure of W50 \citep{blundell2004,roberts2008,blundell2018}.
The jet velocity of SS433 is $v_{\rm jet} = 0.2581c$, where $c$ is the speed of light, and the precession period is 164 days \citep{1984ARA&A..22..507M,1989ApJ...347..448M}.
Additionally, the kinetic energy luminosity of jets is $L_{\rm kin,jet} \sim 10^{39}~{\rm erg~s^{-1}}$ \citep{2005A&A...431..575B}.
The wings are usually associated with the jets because the major axis of W50 is well aligned with jets from SS433.
Polarization observations indicate the existence of a helical magnetic field in the eastern wing \citep{farnes2017,2018PASJ...70...27S,2018Galax...6..137S}.

Non-thermal X-ray and gamma-ray observations indicate that the W50/SS433 system is an efficient site for cosmic-ray acceleration, and the emissions have been detected along the major axis of the jets \citep{1996A&A...312..306B,brinkmann1996,2007A&A...463..611B,1997ApJ...483..868S}.
Several prominent regions have been identified as X-ray hotspots and are labeled e1, e2, and e3 on the eastern side and w1 and w2 on the western side.
In particular, very-high-energy $\gamma$-rays ($> 25~{\rm TeV}$) have been detected from around e1, e2, and w1 hotspots by the High Altitude Water Cherenkov observatory \citep{2018Natur.564E..38A}.
The spectral energy distribution at the hotspots is consistent with the model for the leptonic scenario.
This scenario is that relativistic electrons scatter cosmic microwave background photons to TeV energies via the inverse Compton process.
In addition, no TeV $\gamma$-ray signal from the central binary system has been detected.
The hotspots should therefore be in-situ particle acceleration sites associated with the SS433 jets.
Moreover, several studies provided observational evidence of $\gamma$-ray emissions at MeV to GeV energy levels from not only around X-ray hotspots but also around the spherical shell of W50 \citep[e.g.,][]{2019ApJ...872...25X, 2019A&A...626A.113S}.
Detailed analysis of over 10 years of GeV gamma-ray data recorded by the Fermi Gamma-ray Space Telescope revealed that the precessional period of the SS433 jets is linked with the time variability of $\gamma$-ray excess in the northern shell of W50 \citep{2020NatAs.tmp..158L}.
These observations also strongly support that W50/SS433 has plausible cosmic-ray acceleration sites.
However, it remains unclear whether the outstanding acceleration is due to the interaction between the jet and the supernova remnant (SNR) or just the SS433 jets themselves.

W50 has a unique radio morphology, called the manatee nebula, and the formation mechanism of the W50/SS433 system has been debated for many years.
One scenario is that W50/SS433 formed through interaction between the jets of SS433 and wind bubbles, which were swept up by the stellar wind of the companion star of SS433 \citep{1980ApJ...238..722B,1983MNRAS.205..471K}.
In addition, \cite{2014ApJ...789...79A} suggested that the shell of W50 was formed by winds from an accretion disk with a mass accretion rate exceeding the Eddington limit.
These scenarios pose the problem that it takes a long time for a wind explosion to reach 40--50 pc.
The most popular scenario is that the jets and SNR shell interact.
The circular shell of W50 is the shell of an SNR, resulting from a supernova explosion at SS433's birth.
The interaction between the jets and SNR shell form the elongated two-side wings \footnote{Some papers refer to 'wings' as 'ears'.}.
A structure protruding from a supernova shell is often found in core-collapsed SNRs and looks like the elongated structure of W50 \citep[e.g.,][]{1998MNRAS.299..812G}.
\cite{2017MNRAS.468.1226G} estimated the additional energy used to create the wings of the core-collapse SNR to be 1--10 percent of the explosion energy of the supernova under simple geometrical assumptions.
However, these wings are shorter than the radius of their own SNR.
In contrast, the length of the eastern wing of W50 is much longer than the radius of the shell.
For the wing length to be comparable to the shell radius, the additional kinetic energy needed to extend the W50 wing must be the same as the supernova explosion energy.
Simply assuming the kinetic energy luminosity of the SS433 jets is $10^{39}~{\rm erg~s^{-1}}$ and the explosion energy is $10^{51}~{\rm erg}$, the long-term activity of SS433 for 10--100 kyrs is essential.

The interaction between the SS433 jets and the SNR shell has been studied by conducting hydrodynamic simulations.
\cite{2000A&A...362..780V} conducted the hydrodynamic simulation of jet propagation inside an SNR, as modeled using the analytical Sedov solution.
Additionally, they showed that the cocoon gas, which is heated at the jet's terminal shock, slightly increases the shell expansion velocity near the interaction region.
A three-dimensional simulation conducted by \cite{2008MNRAS.387..839Z} revealed that the exponential density profile of an ambient ISM identified by HI observations reproduced the length ratio of the eastern–western wings.
The reason for the difference in the wing lengths is that a jet propagating into a region of increasing density drastically decelerates.
\cite{2011MNRAS.414.2838G} reported two-dimensional hydrodynamical simulation results for models of several jets propagating in an SNR shell; i.e., precession jets, cone jets, and fireball jets.
They suggested that multi-episodic jet activity is needed for the formation of W50.
Recently, \cite{2019ASSP...55...71M} performed a high-resolution adaptive-mesh refinement hydrodynamical simulation of precessing jets propagating in an SNR.
Their results indicate that precession does not affect the large-scale morphology of W50, while the jets are well collimated at $\sim 0.068~{\rm pc}$ from SS433 and become hollow \citep{2015A&A...574A.143M,2019ASSP...55...71M}.
Note that \cite{2014A&A...561A..30M} reported that precession would be dynamically important from the SS433 scale to the sub-parsec scale because the precessing jets transport more kinetic energy into the ISM than do straight jets owing to the increase in the area of interaction between the jets and ISM.
These hydrodynamical models for the interaction of the jets and SNR can explain the morphology of W50 within a realistic time, which is 10--100~kyrs, in contrast with the wind model.
However, there are problems.
Although the jet termination regions are expected to be sites of efficient particle acceleration in the SNR and jet scenarios, prominent X-ray emissions are not detected in the terminate regions of the wings.
Moreover, there is no physical explanation of where X-ray hotspots form in the intermediate region of SS433 jets.
\cite{2011MNRAS.414.2838G} pointed out differences between simulation results and observations.
In particular, the connections between the jet and SNR shell are a smooth transition in contrast with simulation results (see Figure 10 in \citealt{2011MNRAS.414.2838G}).

In terms of the morphology of continuous jets produced by an active galactic nucleus (AGN), some numerical simulations of light jets, which have a lower density than the surrounding medium, reproduced the structure of the shell and wings \citep{2009MNRAS.400.1785G, 2011ApJ...733...58H,2017A&A...606A..57R}.
When the jet density is lower than the ISM density, we can use the simple formula of the light jet to estimate the advance speed of the jet head and the lateral expansion velocity \citep[e.g.,][]{1982A&A...113..285N,2003A&A...398..113K}.
The important parameter for the advance speed is the mass ratio of the injected jet density to ISM density, $\eta = \rho_{\rm jet}/\rho_{\rm ISM}$.
The advance speed of light jets $v_{\rm head}$ can be determined analytically from the momentum balance at the contact discontinuity between the jets and ISM. The advance speed is approximately \citep{2002PhDT.........2K}
\begin{equation}
    \label{eq:vhead}
  v_{\rm head} = \frac{\sqrt{A\eta}}{1+\sqrt{A\eta}} v_{\rm beam},
\end{equation}
where $A$ and $v_{\rm beam}$ are respectively the ratio of the beam to head area and the beam velocity.
Equation \ref{eq:vhead} indicates that the advance speed also depends on the ratio of the beam to head.
The ratio of the beam to head is close to unity for a heavy jet ($\eta \sim 1$) and decreases to 0.1 for a light jet ($\eta \ll 1$) \citep{2002PhDT.........2K}.
Thus, light jets decelerate rapidly, and the propagation enters a non-linear stage.
Meanwhile, the lateral expansion of light jets is well described by the blast-wave equation of motion after the bowshock reachs at $\sim R_{\rm jet}/2\eta^{0.25}$, where $R_{\rm jet}$ is the jet beam radius. \citep{2003A&A...398..113K,2005A&A...431...45K}.
Light jets have strong backflows whose plasma continuously goes toward the core, and the jets become supersonic expansion in the lateral direction.
When we simply assume constant energy injection $E = L_{\rm kin,jet} t$ and the density of the homogeneous ISM $\rho_{\rm ISM}$, the radius of lateral expansion can be written as
\begin{equation}
  \label{eq:blast}
  R_{\rm lateral} = \left( \frac{5L_{\rm kin, jet}t^3}{4\pi\rho_{\rm ISM} }\right)^{1/5} = \left( \frac{5}{4} R^2_{\rm jet}\eta v_{\rm beam}^3 t^3  \right)^{1/5},
\end{equation}
where $L_{\rm kin,jet} = \pi R^2_{\rm jet} \rho_{\rm jet} v^3_{\rm beam}$ is the kinetic energy luminosity of the jets.
Therefore, $v_{\rm beam}, R_{\rm jet}$, and $\eta$ are quantities important to the morphology of light jets.

In this paper, we propose new modeling of the magnetohydrodynamics (MHD) of W50/SS433 generated by jets and backflows.
The jets injected during the period 20--90 kyrs are two-sided, light, supersonic, and magnetized.
We conduct four simulations for jets with different radii and density ratios, and we present a mechanism for the formation of a shell and wings like those of W50.
The remainder of the paper is structured as follows.
We describe the basic equations and numerical setup of the jets and ISM in Section \ref{sec:setup}.
In section \ref{sec:results}, we present our four models for MHD simulations.
We discuss the results and other formation scenarios of W50/SS433 in Section \ref{sec:discussion} and give a summary in Section \ref{sec:summary}.
\section{Numerical Setup} \label{sec:setup}
We solve ideal MHD equations in the axisymmetric cylindrical coordinate system $(R,\phi,z)$.
The equations are
\footnotesize
\begin{align}
 &\frac{\partial \rho}{\partial t} + \nabla \cdot (\rho \bm{v}) = 0, \\
 &\rho \left( \frac{\partial \rho}{\partial t} + \bm{v} \cdot \nabla \bm{v} \right) = -\nabla \left( p + \frac{B^2}{8\pi} \right) + \frac{1}{4\pi} (\bm{B} \cdot \nabla) \bm{B}, \\
 &\frac{\partial}{\partial t} \left( e + \frac{\rho v^2}{2}+ \frac{B^2}{8\pi} \right) + \nabla \cdot \left[ (e+p+ \frac{\rho v^2}{2})\bm{v} - \frac{(\bm{v}\times \bm{B}) \times \bm{B}}{4\pi}\right] = 0, \\
 &\frac{\partial \bm{B}}{\partial t} = \nabla \times (\bm{v}\times \bm{B}), \\
 &\nabla \cdot \bm{B} = 0, \\
 &\frac{\partial f}{\partial t} + \nabla \cdot (f \bm{v}) = 0,
 \label{eq:induction}
 \end{align}
 \normalsize
 where $\rho, \bm{v}, p, \bm{B}, e,$ and $f$ are respectively the density, velocity, pressure, magnetic field, internal energy density, and jet tracer function.
To distinguish the jets and ISM, the jet tracer function is set equal to unity for the injected jet flow and equal to zero for the ISM.
We adopt the ideal gas law and give the internal energy density as $e = (\gamma - 1)p$, where $\gamma=5/3$ is an adiabatic index.
The plasma composition is 70 percent hydrogen and 30 percent helium by mass so that the mean molecular weight is 1.3.
We use the MHD code CANS+ \citep{2019PASJ...71...83M}.
The code is based on the HLLD approximate Riemann solver \citep{2005JCoPh.208..315M}. The code implements fifth-order accuracy in space and third-order accuracy in time.
We adopt the hyperbolic divergence cleaning method to reduce numerical errors under the divergence-free condition of the magnetic field \citep{2002JCoPh.175..645D}.
The number of cells in our simulation box is $(N_{\rm R},N_{\rm z})=(1200,4000)$.
The physical dimensions of the box are $0 < R < 60~{\rm pc}$ and $-80 < z < 120~{\rm pc}$.
We thus adopt a uniform grid in both directions, $\Delta R = \Delta z = 0.05~{\rm pc}$.
We apply a symmetric boundary condition at $R=0$ and an outflow boundary condition, which sets a zero gradient across the boundary, at other boundaries.

In this work, we assume the distance to SS433 as 5.5 kpc \citep{2007MNRAS.381..881L}.
The ISM around SS433 thus seems to have a galactic exponential profile \citep{1998MNRAS.294..429D}.
We adopt a modified exponential density profile for the initial density of the ISM:
\begin{align}
 \rho_\mathrm{amb} =
 \begin{cases}
  \mu m_\mathrm{p} n_0 & (z \geq -20) \\
  \mu m_\mathrm{p} n_0 \exp \left(-\dfrac{R}{R_\mathrm{d}} - \dfrac{z + 20}{Z_\mathrm{d}} \right) & (z < -20),
 \end{cases} \label{eq:HI_gas}
\end{align}
where $R_{\rm d}=5.4~{\rm kpc}$ is the scale length of the disc, $Z_{\rm d}=30~{\rm pc}$ is the scale height of the galactic disc, $n_0 = 0.1 {\rm~cm^{-3}}$, $m_{\rm p}$ is the proton mass, and SS433 is located at $z=0$ and $R=0$.
Note that the HI intensity map of the GALFA-HI survey shows that the constant density distribution of the ISM is a reasonable assumption in the eastern region of SS433 \citep{2011ApJS..194...20P}.
The western and eastern directions respectively correspond to $z<0$ and $z>0$ directions in this simulation.
The ISM is in pressure equilibrium, and the temperature of the ISM is $10^4~{\rm K}$ at the location of SS433.
We assume that the uniform magnetic field is parallel to the z-axis and that the plasma $\beta\equiv p/(B^2/8\pi)$ is 10 ($B_z \sim 0.59~\mu$G).

The two-side jets are injected by a cylindrical nozzle along the z-axis at the origin.
We adopt a velocity $v_{\rm jet} = 7.8 \times 10^9~{\rm cm~s^{-1}}$ and initial Mach number ${\mathcal M} = 10 $ for the jets in all runs.
We make calculations for four models, which have different density contrasts $\eta = 10^{-2}$ and $10^{-3}$ and different jet radii $R_{\rm jet} = 1$ and $3.33$~pc.
The jet nozzle length is 2 pc ($|z| < 1$ pc).
The jet kinetic energy luminosity is thus written as
\begin{eqnarray}
L_{\rm kin} &=& \pi R^2_{\rm jet} \rho_{\rm jet}v_{\rm jet}^3 = \pi R^2_{\rm jet} \eta \rho_{\rm 0}v_{\rm jet}^3 \\ \nonumber
 &\sim& 3.1 \times 10^{42} \eta \left( \frac{R_{\rm jet}}{1~{\rm pc}}\right)~{\rm erg~s^{-1}}.
\end{eqnarray}
We inject a toroidal field $B_{\phi} = B_{\rm jet}sgn(z)\sin^4(\pi R/R_{\rm jet})$, where $sgn$ is the sign function, in the jet nozzle.
We set $\beta = 1$ at $R=0.5 R_{\rm jet}$ and hence $B_{\rm jet} = 140~\mu$G for all runs.
The parameters are summarized in Table \ref{tab:common} and \ref{tab:model}.
\begin{table}
  \begin{center}
  \caption{Common physical parameters in our runs}
  \begin{tabular}{ccc} \hline
      ISM density at the origin     & $\rho_{\rm 0}$           & $2.17\times10^{-25} ~{\rm g~cm^{-3}}$  \\
      ISM temperature at the origin & $T_{\rm 0}$              & $10^4~{\rm K}$\\
      ISM magnetic field            & $B_{z,\rm ISM}$          &   0.59 $\mu$G   \\ \hline
      Jet velocity                  & $v_{\rm jet}$            &  $7.8 \times 10^9 {\rm cm~s^{-1}}$  \\
      Jet sonic Mach number         & ${\mathcal M_{\rm jet}}$ & 10     \\
      Jet plasma $\beta$            & $\beta_{\rm jet}$        &  1   \\ \hline
    \end{tabular}
  \end{center}
  \label{tab:common}
\end{table}
\begin{table}
  \begin{center}
  \caption{Parameters in runs for different jets }
  \begin{tabular}{cccc} \hline
        Run & $\eta (=\rho_\mathrm{jet}/\rho_0)$ & $R_{\rm jet}$ [pc] & $L_{\rm kin} [{\rm erg~s^{-1}}]$ \\ \hline
        H-R3   & $10^{-2}$ & 3.33 & $3.1\times 10^{41}$         \\
        H-R1   & $10^{-2}$ & 1    & $3.1\times 10^{40}$        \\
        L-R3   & $10^{-3}$ & 3.33 & $3.1\times 10^{40}$         \\
        L-R1   & $10^{-3}$ & 1    & $3.1\times 10^{39}$        \\ \hline
    \end{tabular}
    \end{center}
    \label{tab:model}
\end{table}

\section{Numerical Results} \label{sec:results}
\subsection{Morphology}
\begin{figure*}
     \begin{center}
       \includegraphics[width=1.0\textwidth,bb = 0 0 461 346]{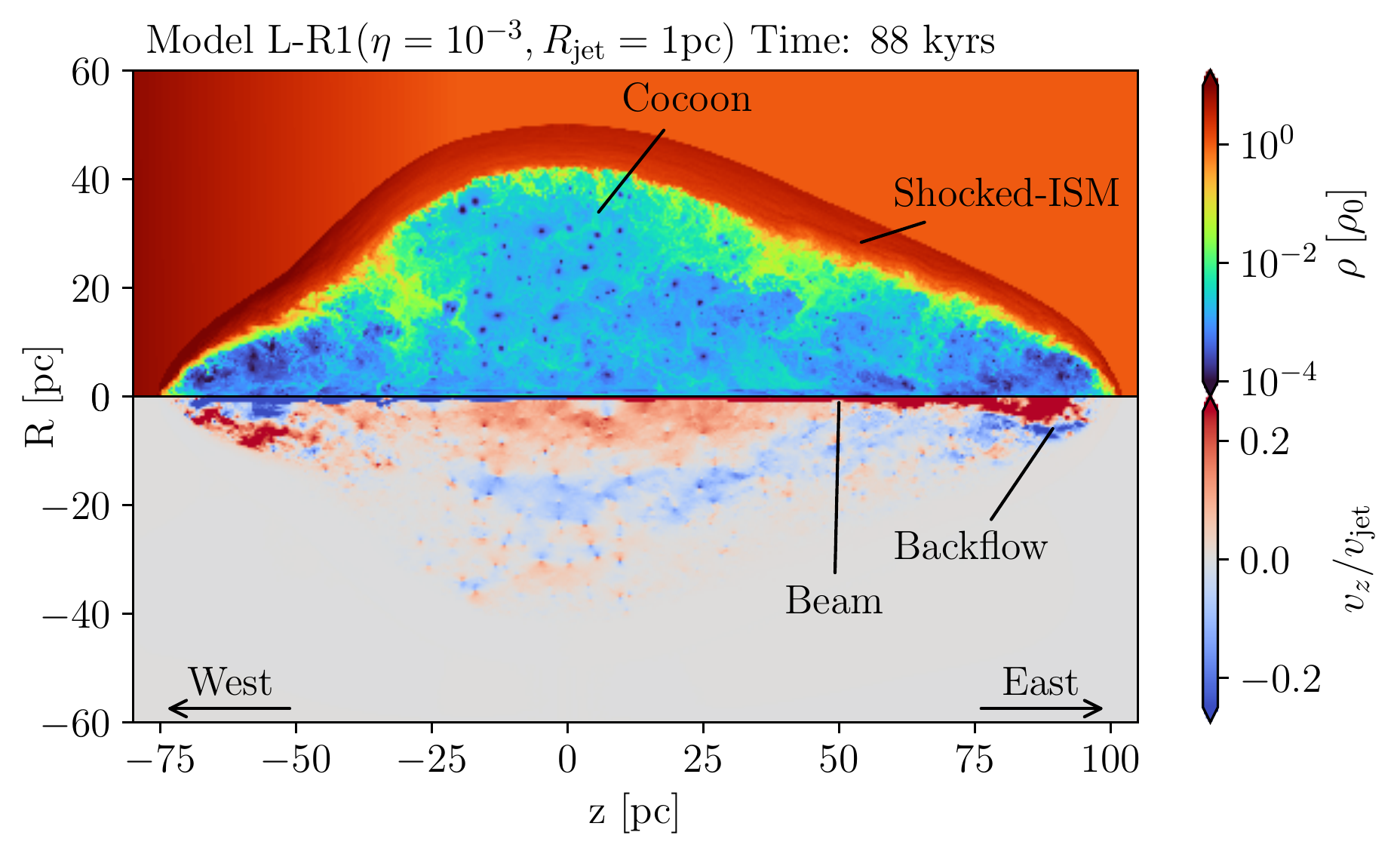}
       \caption{Density ({\it top}) and z-direction velocity ({\it bottom}) maps for run L-R1 at $t = 88~{\rm kyrs}$. }
       \label{fig:l-r1_rovz}
     \end{center}
\end{figure*}

The basic structure in all runs has a dynamic behavior similar to that found in previous light-jet simulations \citep[e.g.,][]{2020MNRAS.493.5761O}.
Therefore, we briefly summarize the basic structure of light jets and then describe the overall morphology of each run in detail.
Figure \ref{fig:l-r1_rovz} shows counters of the gas density ({\it top}) and z-direction velocity ({\it bottom}) for run L-R1 at $t = 88~{\rm kyrs}$.
The basic structures of supersonic jets, such as a beam, backflow, cocoon, and thick shell (shocked ISM) are seen.
Surrounding plasma mixes into the cocoon owing to Kelvin--Helmholtz (KH) instabilities at the contact discontinuity between the jet gas and shocked ISM.
In particular, backflows from the tips of the jet interact with each other, and vortex motions are highly developed in the cocoon.
The background density spatially increases at $z < -20$ pc in the western direction, and the advance of the jet in the western direction thus slows appreciably.
The magnetic field distribution and its detail are reported in section \ref{subsec:mag}.

\begin{figure*}
    \begin{center}
    \includegraphics[width=1.0\textwidth,bb = 0 0 461 346]{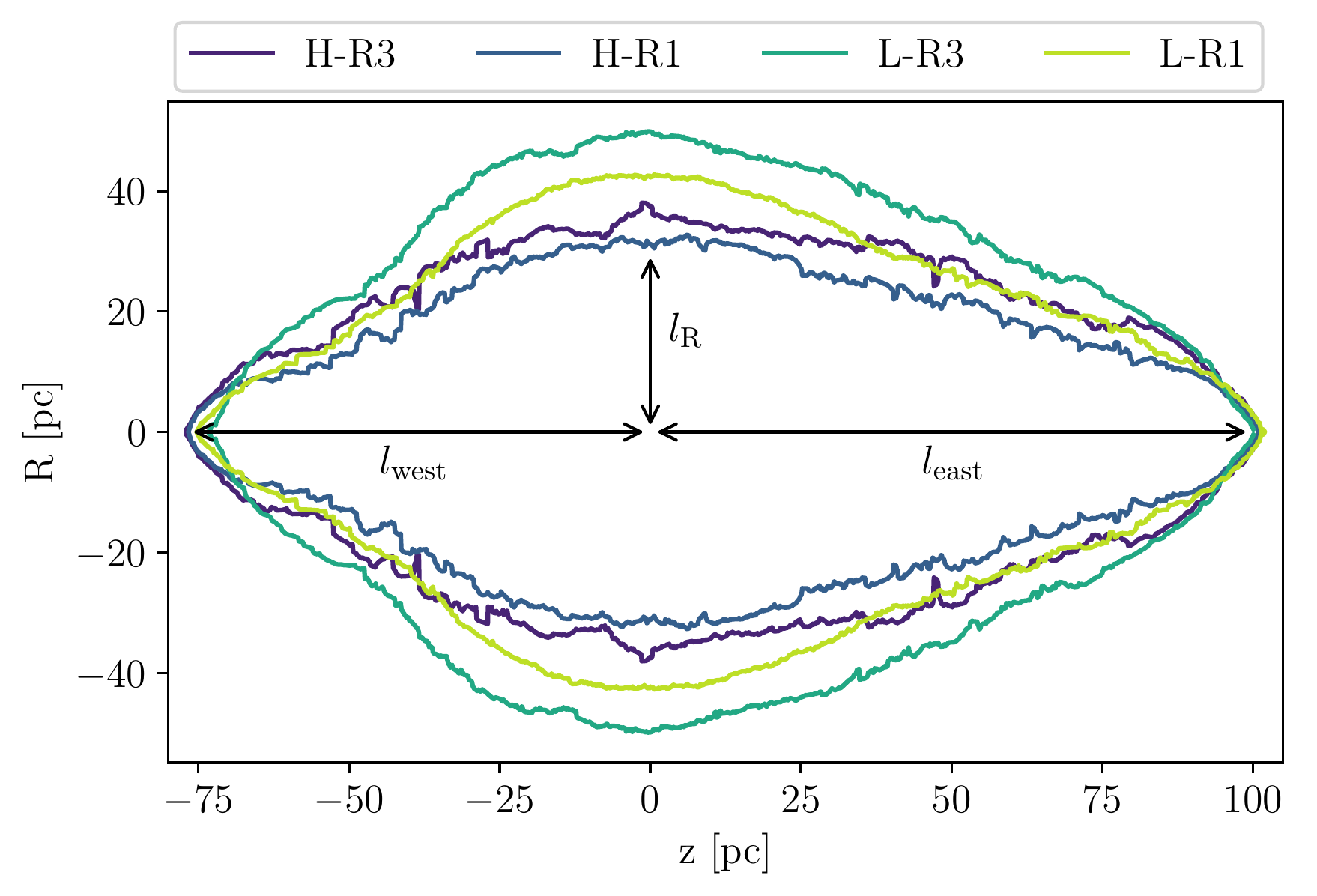}
    \caption{Shapes of jets for all runs. Runs H-R3, H-R1, L-R3, and L-R1 are at $t = 16, 29, 55,$ and $88~{\rm kyrs}$, respectively. We denote the shell radius and the lengths from the origin to the eastern/western edges of the jet as $l_{\rm R}$, $l_{\rm east}$, and $l_{\rm west}$.
    }
    \label{fig:outline}
    \end{center}
\end{figure*}

Figure \ref{fig:outline} shows the shape of the jet for all runs when the contact discontinuity between the jet and ISM reaches 100 pc.
Because the advance speed of jets depends on the injection condition, the times at which the jets reaches 100 pc are different; i.e., H-R3, H-R1, L-R3, and L-R1 are at $t=16, 29, 55,$ and $88~{\rm kyrs}$ respectively.
We here define the outer shape of jets as the contact discontinuity rather than a bow shock.
However, it is difficult to identify the contact discontinuity between the jet plasma and shocked ISM because the gases are highly mixed by the KH instability.
We therefore define the outer shape as the maximum radius of each $z$ at which the jet tracer function $f$ is less than $10^{-5}$.
An analogous definition has been widely used in previous works on jet propagation \citep[e.g.,][]{2009MNRAS.400.1785G}.
We here denote the shell radius and the lengths from the origin to the eastern/western edges of the jet as $l_{\rm R}$, $l_{\rm east}$, and $l_{\rm west}$.

We find that all runs have a circular shell centered on the origin when jets reach $\sim 100$ pc.
The ratio of $l_{\rm east}$ to $l_{\rm west}$ is $100:75$ for all runs.
Lighter jets such as L-R1 and L-R3 form a larger shell.
A lower density ratio corresponds to a lower lateral expansion velocity (see equation \ref{eq:blast}).
However, because the kinetic energy luminosity of light jets is less than that of heavy jets, the advance of the light jet becomes much slower than that of the heavy jets.
Lighter jets thus have enough time to form a larger circular shell while they propagate a distance of 100 pc.
Run L-R1 therefore reproduces the characteristic morphology of W50, including a circular shell and elongated wings.
Note that the wings of L-R3 are much wider than those of W50.
\begin{figure}
  \begin{center}
  \includegraphics[width=1.0\columnwidth,bb = 0 0 461 346]{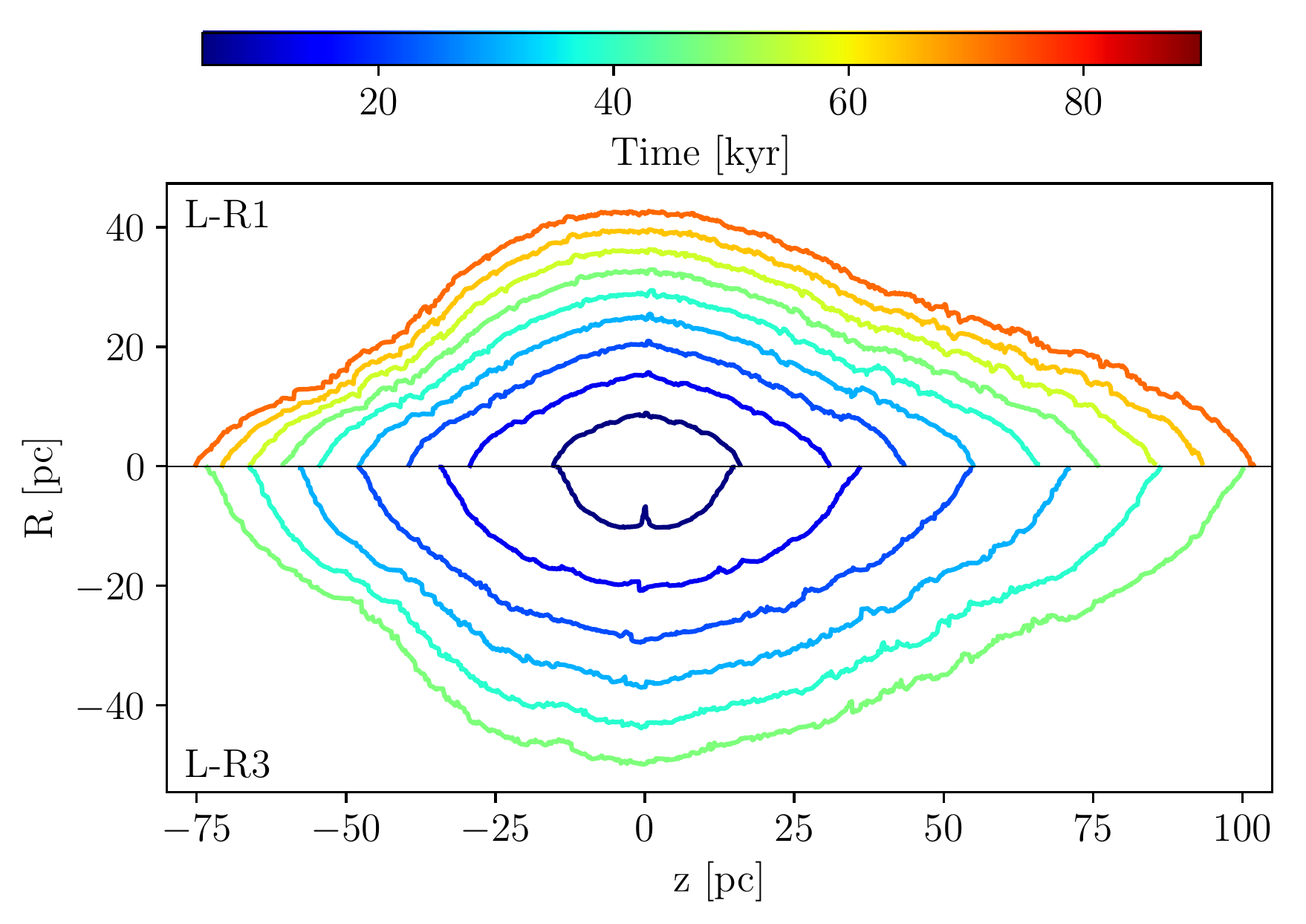}
  \caption{Time evolution of the shape of jets of runs L-R1 ({\it top}) and L-R3 ({\it bottom}).
  Colors denote the shape at $t = 8, 18, 28, 38, 48, 58, 68, 78, 88$ kyrs for run L-R1 and $t = 5, 15, 25, 35, 45, 55$ kyrs for run L-R3.}
  \label{fig:outline_tevo}
  \end{center}
\end{figure}

\subsection{Time evolution}  \label{sec:results_tevo}
\begin{figure*}
  \begin{minipage}{0.5\hsize}
    \includegraphics[width=0.95\textwidth,bb=0 0 461 346]{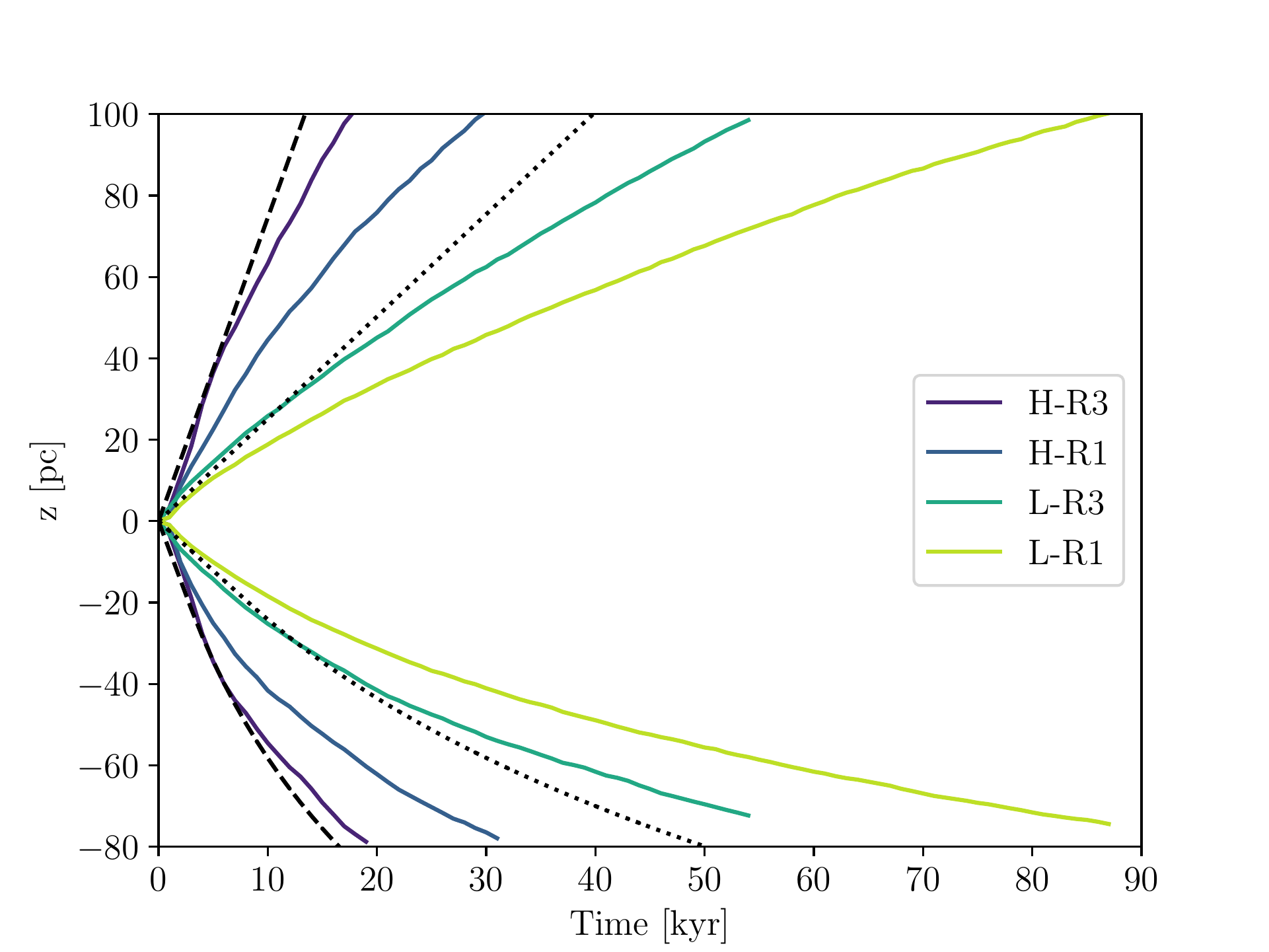}
  \end{minipage}
  \begin{minipage}{0.5\hsize}
    \includegraphics[width=0.95\textwidth,bb=0 0 461 346]{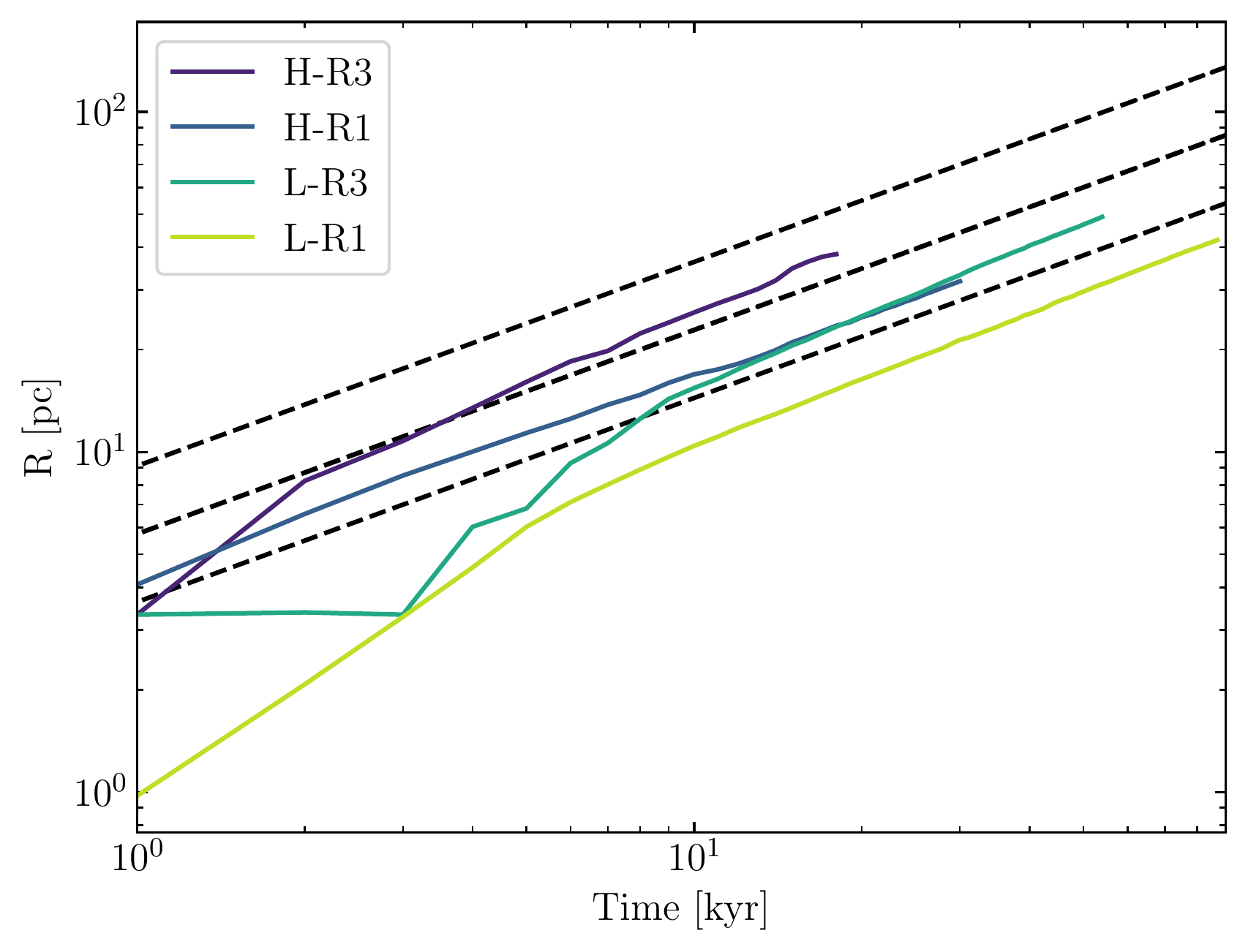}
  \end{minipage}
  \caption{Positions of the tips of the eastern/western jets ({\it left}) and the shell radius ({\it right}) as a function of time for each run.
  Dashed and dotted lines in the left panel show analytic results of equation \ref{eq:vhead} and \ref{eq:vhead_exp} for runs H and L, respectively.
  Dashed lines in the right panel show analytic results obtained using equation \ref{eq:blast}.
  }
  \label{fig:zpos}
\end{figure*}
\begin{figure*}
  \begin{minipage}{0.5\hsize}
    \includegraphics[width=0.95\textwidth,bb=0 0 413 317]{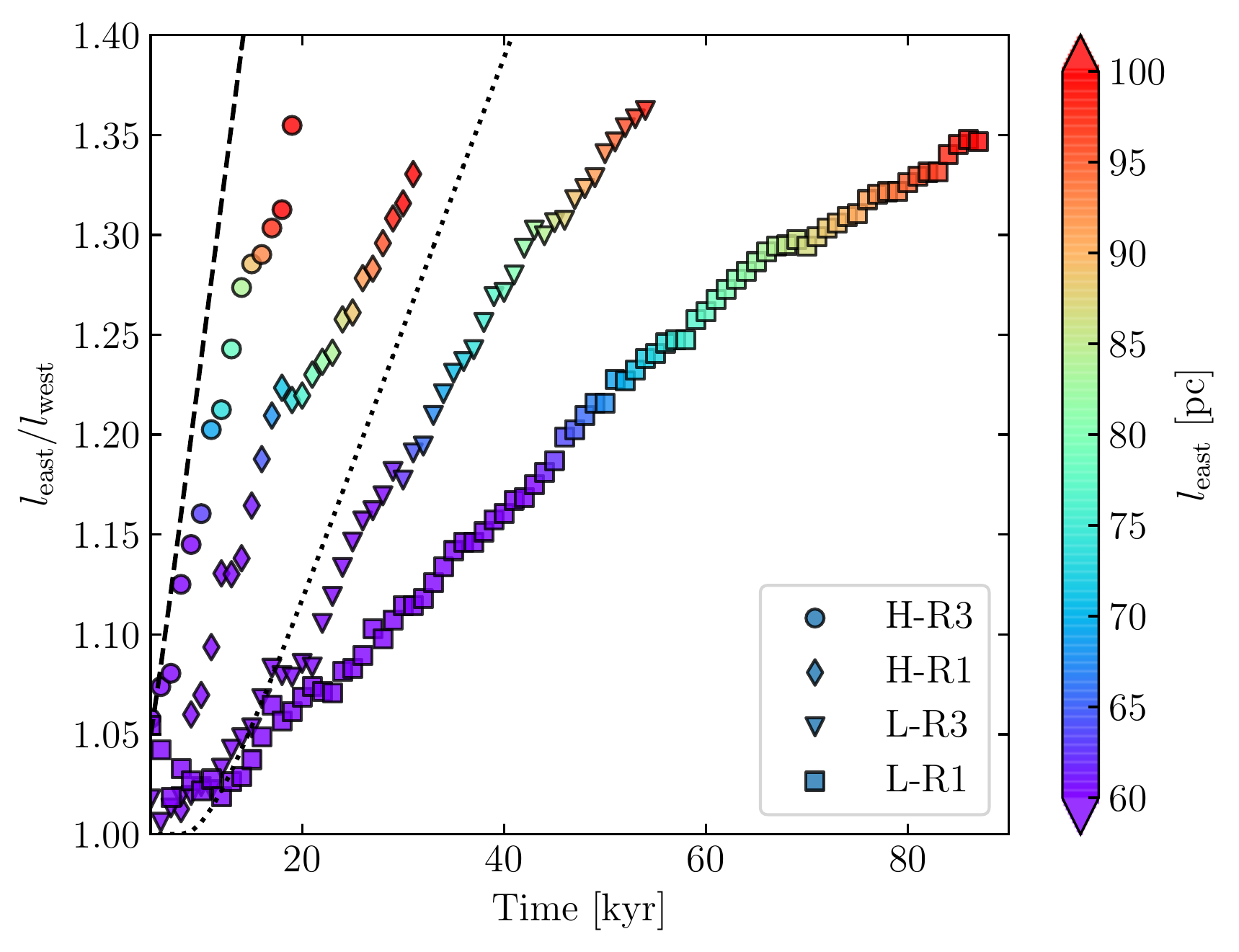}
  \end{minipage}
  \begin{minipage}{0.5\hsize}
    \includegraphics[width=0.95\textwidth,bb=0 0 461 346]{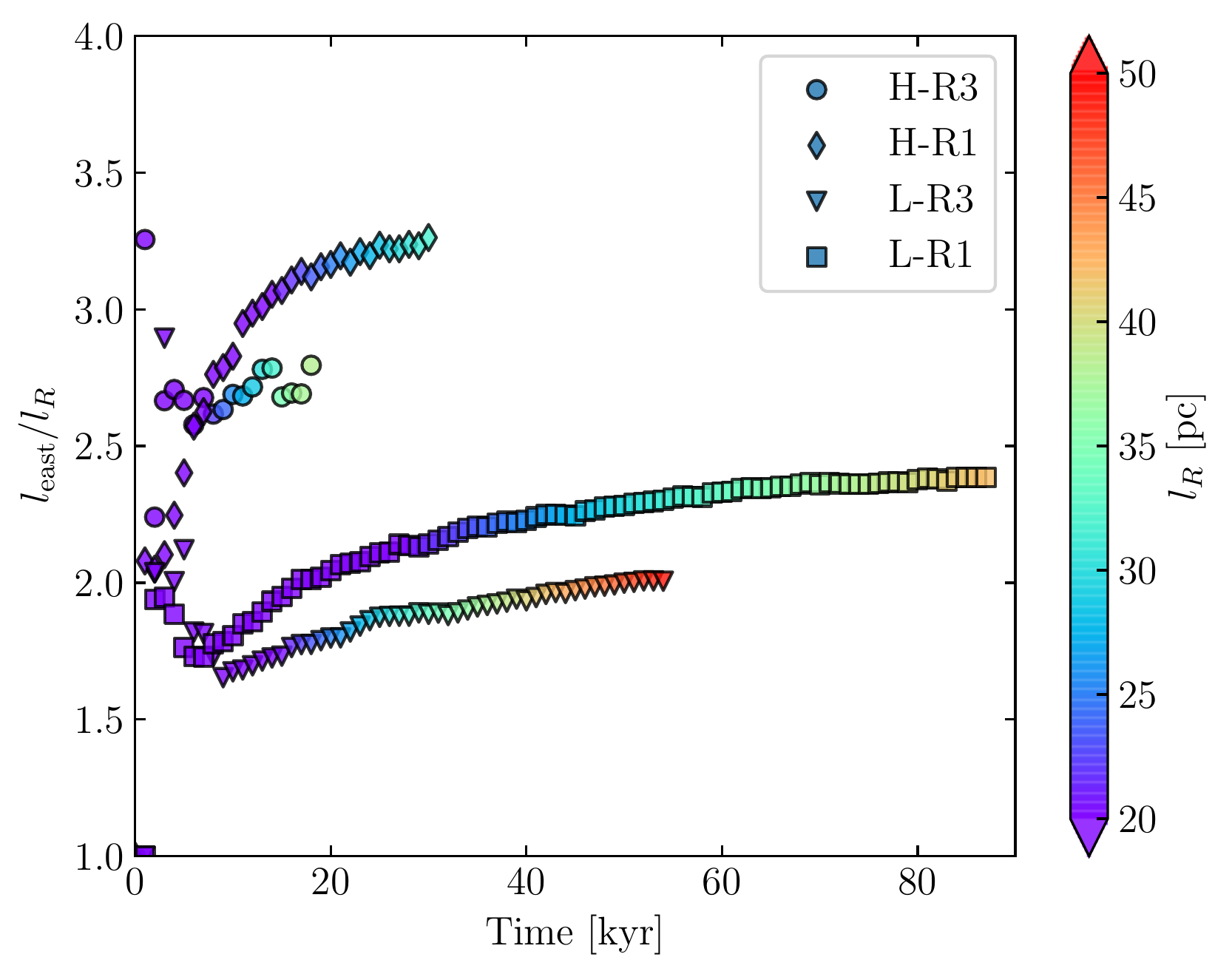}
  \end{minipage}
  \caption{ $l_{\rm east}/l_{\rm west}$ ({\it left}) and $l_{\rm east}/l_{\rm R}$ ({\it right}) as functions of time for each run.
  Symbol colors denote the position of the tip of the eastern jet ({\it left}) and the shell radius ({\it right}).
  Dashed and dotted lines in the left figure show analytic values obtained using equations \ref{eq:vhead} and \ref{eq:vhead_exp} for runs H and L, respectively.
  }
  \label{fig:length_ratio}
\end{figure*}

In this section, we describe the time evolution of the jet morphology (Figure \ref{fig:outline_tevo}).
Top and bottom panels show L-R1 and L-R3, respectively.
Colors denote the shapes at $t = 8, 18, 28, 38, 48, 58, 68, 78, 88$ kyrs for run L-R1 and $t = 5, 15, 25, 35, 45, 55$ kyrs for run L-R3.
In the early period of $t \lesssim 30$ kyrs, the shapes of L-R1 and L-R3 are spheroids rather than circular shells.
The spheroidal shape is well known from previous simulations and is similar to that of FR II radio sources, such as 3C 452 \citep[e.g.,][]{2015MNRAS.454.3403H}.
We find that a circular shell and wings then form.
The region of the wing is filled with back-flowing plasma and it is thus difficult for the region to expand rapidly like the shell.
Once the shell and wings have formed, they expand as self-similar evolution.

We next consider what physical parameters determine the advance speed of the jet.
We easily estimate the advance speed of the jet in the positive z direction adopting equation \ref{eq:vhead}.
Meanwhile, the density distribution in the western direction exponentially increases, and the gradient effect should thus be included in the expression for the advance speed:
\begin{equation}
  v_{\rm head} = \frac{dz}{dt} = \frac{ \sqrt{\eta(z)} }{ 1 + \sqrt{ \eta(z)} } v_{\rm beam},~~~ \eta(z) = \eta_0 \exp{ \left( \frac{z-20}{Z_{\rmf d}} \right)}.
  \label{eq:vhead_exp}
\end{equation}
Here, we simply assume that $A = 1$ and $\rho_0, v_{\rm beam}$ are constant values.

Figure \ref{fig:zpos} ({\it left}) shows the positions of the tips of the eastern/western jets as a function of the time for each run ({\it solid lines}) and analytic lines ({\it dashed lines} $\eta=10^{-2}$, {\it dotted lines} $\eta=10^{-3}$) of equation \ref{eq:vhead} ($z > 0$) and equation \ref{eq:vhead_exp} ($z < 0$, towards to the galactic plane).
All of the lines at $z > 0$ (eastern) are not in good agreement with analytic values because the advance speed decreases rapidly.
The area of the jet head increases during the propagation of the jets; i.e., $A$ is decreasing in equation \ref{eq:vhead} because vortices grow downstream of the terminal shock.
The jets, which have a large radius, have a small deceleration rate for both runs H and L because $A$ remains at unity for a longer time than in the case of jets having a small radius.
When a jet propagates toward the galactic plane (in the western direction), which means the ambient density increases as the distance increases ($z<0$), equation \ref{eq:vhead_exp} well describes the results of runs H-R3 and L-R3.

We next describe the results of the time evolution of radial expansion for each run.
Figure \ref{fig:zpos} ({\it right}) shows the shell radius $l_{\rm R}$ as a function of time for each run ({\it solid lines}).
Dashed lines are the time evolution of the analytical model (equation \ref{eq:blast}) whose kinetic energy luminosity is $L_{\rm kin,jet}~= 10^{39}, ~10^{40},$ and $10^{41}~{\rm erg~s^{-1}}$.
All runs have $l_{\rm R} \propto t^{2/5}$ in accordance with the analytical solution.
Note that equation \ref{eq:blast} describes the length of the bow shock along the $R$ axis instead of the contact discontinuity.
$l_{\rm R}$ is defined by the position of the contact discontinuity, and the expansion rate in all runs is thus lower than that for equation \ref{eq:blast}.
Note that the approximation for equation \ref{eq:blast} is not valid when the density ratio of the ISM to jets is close to unity.
Furthermore, jets with larger radii have more kinetic energy luminosity, and the speed of radial expansion thus depends on the jet's radius.

To discuss the formation mechanism of shells and cocoon, we focus on the length ratio of the two-side jets and the length ratio of the jets and shell radius.
In this work, the shell radius of W50 and the lengths from SS433 to the tips of eastern and western wings are respectively $R_{\rm W50} \sim 48~{\rm pc}$, $L_{\rm east} \sim~121.5~{\rm pc}$, and $L_{\rm west} \sim 86.5~{\rm pc}$ according to radio observations \citep{dubner1998,2011MNRAS.414.2838G}.
Thus, $L_{\rm east}/R_{\rm W50}$ and $L_{\rm east}/L_{\rm west}$ are 2.5 and 1.4, respectively.
Figure \ref{fig:length_ratio} ({\it left}) shows the time evolution of the length ratio $l_{\rm east}/l_{\rm west}$ for all runs.
The dashed and dotted lines are analytic solutions obtained when $\eta = 10^{-2}$ and $10^{-3}$.
The numerical results for all runs and analytic solutions show a linear evolution over time.
In this work, we perform calculations until the tip of one side jet reaches 100 pc, but the length ratio, $l_{\rm east}/l_{\rm west}$, increases with simulation time.
It is thus expected that the length ratio becomes 1.4 when $l_{\rm east} = 120~{\rm pc}$, which is consistent with radio observations.
Interestingly, the length ratios are roughly the same for all runs when the positions of jets are fixed.
These results indicate that $l_{\rm east}/l_{\rm west}$ is determined only by the density profile of the ISM.

We next investigate the time evolution of the ratio of length from the center to the tip of the jet propagation in the eastern direction to the shell radius $l_{\rm east}/l_{\rm R}$ (right panel of Figure \ref{fig:length_ratio}).
For all runs, $l_{\rm east}/l_{\rm R}$ saturates at certain values.
The lower-density runs of L-R1 and L-R3 have a constant value of $l_{\rm east}/l_{\rm R}$ for a long time.
These results mean that outer shapes seem to undergo self-similar expansion in the later phase (see Figure \ref{fig:outline_tevo}).
The ratio of $l_{\rm east}$ and $l_{\rm R}$ strongly depends on $\eta$.
Although the advance speed of the jet head is proportional to $\eta$, the speed of radial expansion does not depend on $\eta$.
Although the advance speed is strongly dependent on $\eta$, the speed of radial expansion is weakly dependent on $\eta$.
Therefore, $l_{\rm east}/l_{\rm R}$ is greater when $\eta$ is closer to unity.
The ratio of the eastern wing to the shell radius of W50/SS433 is $L_{\rm east}/R_{\rm W50} \sim  2.5$, which is consistent with the L-R1 run.
In this work, we assume axisymmetric coordinates.
We note that the advance speed in three-dimensional simulations can be higher than that expected from axisymmetric simulations \citep{2019MNRAS.482.3718P}.
\subsection{Flow and Magnetic Fields} \label{subsec:mag}
\begin{figure}
  \begin{center}
    \includegraphics[width=0.95\columnwidth, bb = 0 0 460 345]{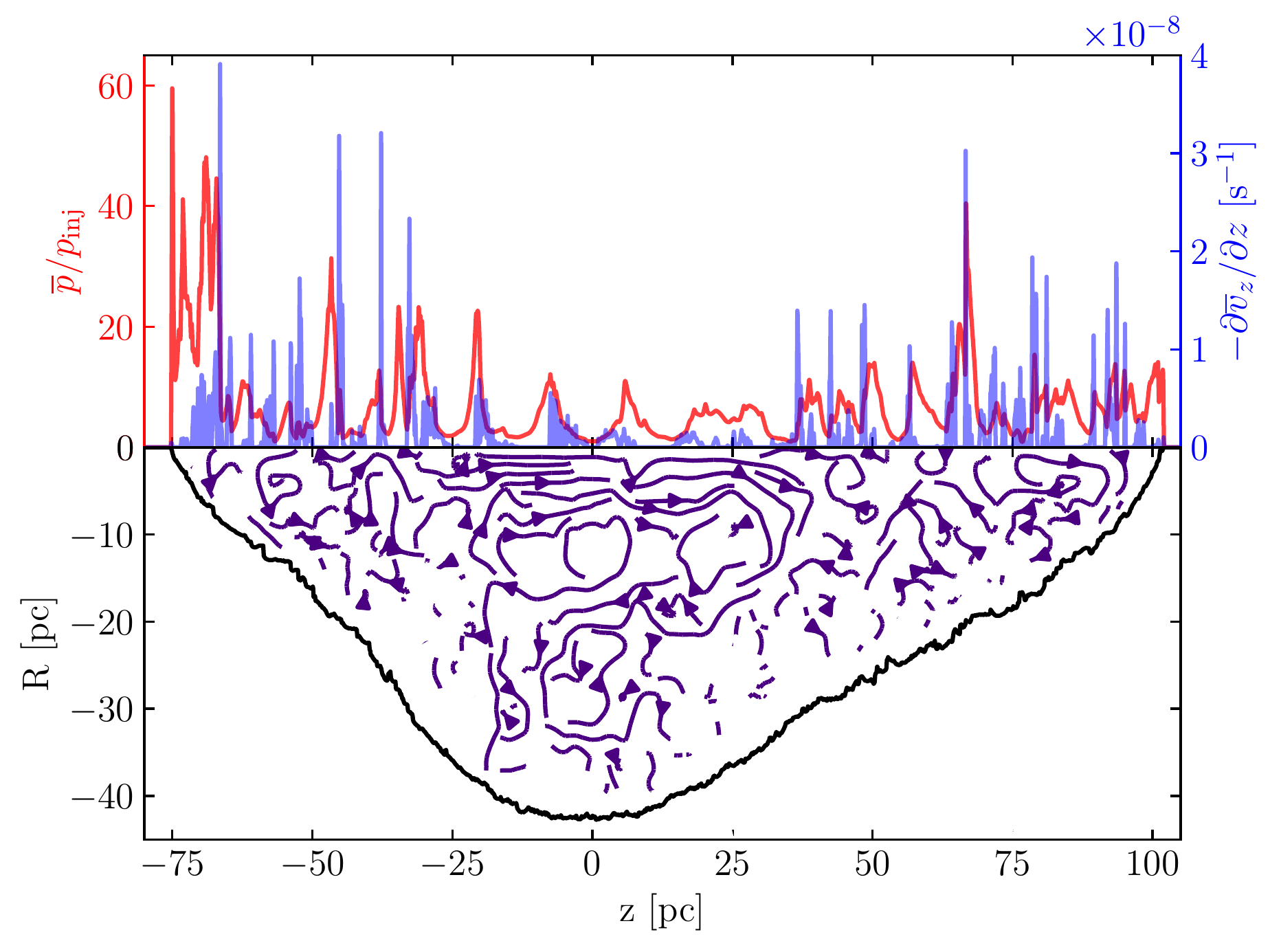}
    \caption{({\it top}) Cut along the z-axis of beam-averaged values of $-\partial \bar{v}_z/\partial z$ ({\rm blue}) and $\bar{p}/p_{\rm inj}$ ({\rm red}) for run L-R1 at $t = 88$ kyrs.
    ({\it bottom}) Velocity stream lines and the outer shape are illustrated in purple and black for run L-R1 at $t = 88$ kyrs.
    }
    \label{fig:shock}
  \end{center}
\end{figure}

Figure \ref{fig:shock} ({\it top}) presents the distributions of physical quantities on the jet axis that show the characteristics of shock waves.
The red and blue curves present the pressure $\bar{p}$ and velocity gradients ($-\partial \bar{v}_z/\partial z$) averaged according to
\begin{equation}
     \bar{f} = \frac{1}{\pi R^2_{\rm jet}} \int^{R_{\rm jet}}_{0} 2\pi R f dR.
\end{equation}
The velocity gradient in the direction of propagation roughly represents the magnitude of the shock wave.
The bottom panel of Figure \ref{fig:shock} shows stream lines of the average velocity.
The velocity gradient (blue curve) has many sharp spikes while the pressure distribution has wide-tailed peaks.
The two are generally correlated and the peaks are considered shock waves, such as reconfinement shocks, oblique shocks, and terminal shocks.
As an example, we see reconfinement shocks at z $\sim$ $-10$, 10, and 20 pc.
After the jet propagates far away, it is difficult to identify the origin of each peak because many reflection waves propagate in various directions.
However, when including the density and velocity distributions (Figure \ref{fig:l-r1_rovz}), we can identify the terminal shocks ($z~\sim$ $-70$, 90 pc) and oblique shocks ($z~\sim$ $-45$, 60 pc) in both side jets.
A terminal shock is a location of energy exchange from jet kinetic energy to thermal energy and it produces backflows.
When the propagation speed decreases, the backflow becomes turbulent and has circular and arc-like motion (see the bottom panel of Figure \ref{fig:shock}).
The backflow therefore interacts with the beam itself and drives the strong oblique shock \citep{2004ApJ...606..804M}.
We estimate the size of the backflow vortex in section \ref{sec:backflow}.
In our calculation, the oblique shock in the eastern jet releases more kinetic energy than the terminal shock.
Meanwhile, the western jet shows the opposite correlation.
The background ISM density distribution increases in the western direction, and the conversion efficiency of the jet kinetic energy to thermal energy at the terminal shock in the western direction is therefore higher than that in the eastern direction.
The red curve shows that the pressure maximum of the jet head at $z = -75$ pc is 3 times that of the jet in the eastern direction ($z = 100$ pc).

\begin{figure*}
  \begin{minipage}{0.6\hsize}
    \includegraphics[width=0.95\textwidth,bb=0 0 461 346]{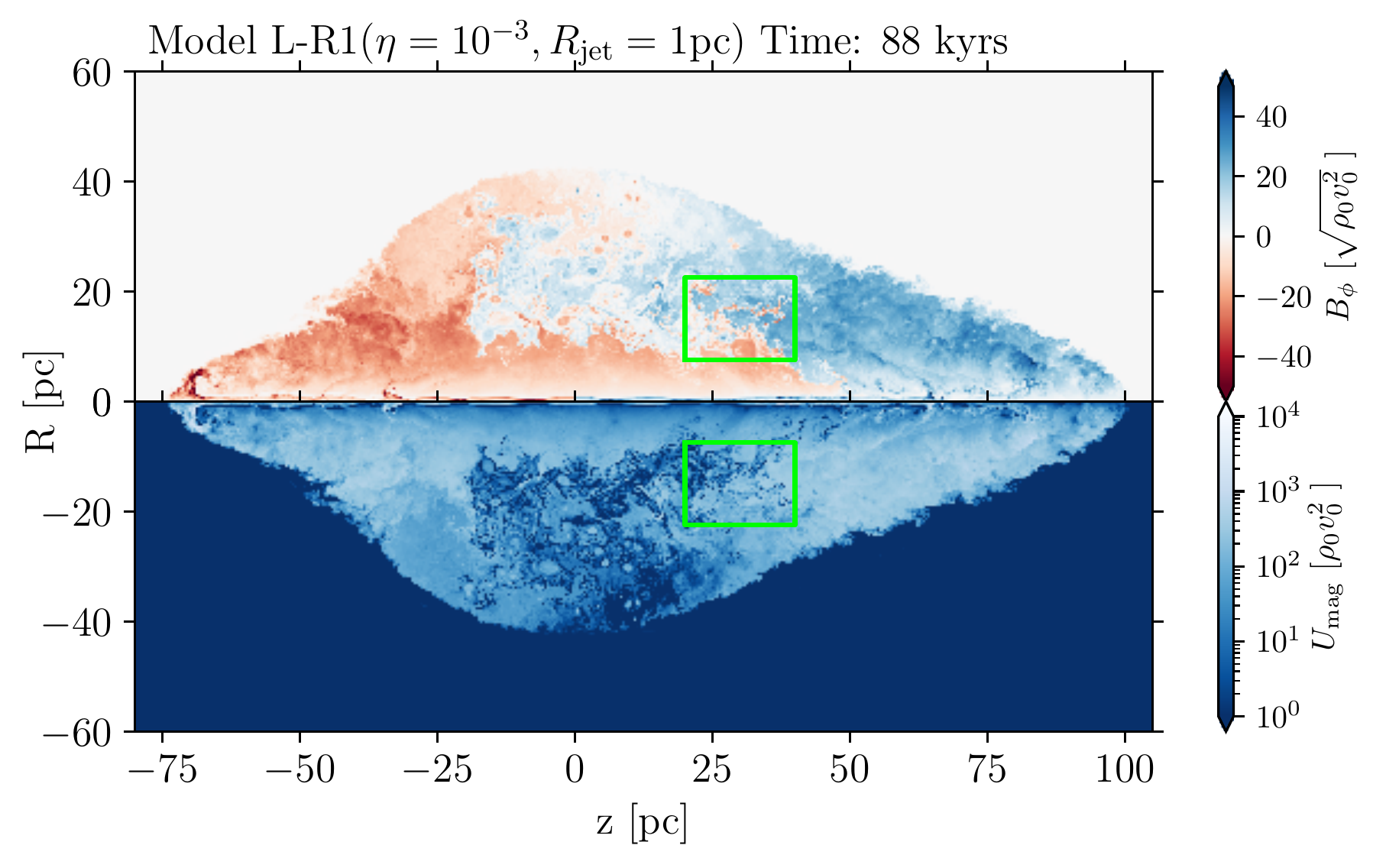}
      \end{minipage}
  \begin{minipage}{0.4\hsize}
    \includegraphics[width=0.95\textwidth,bb=0 0 461 346]{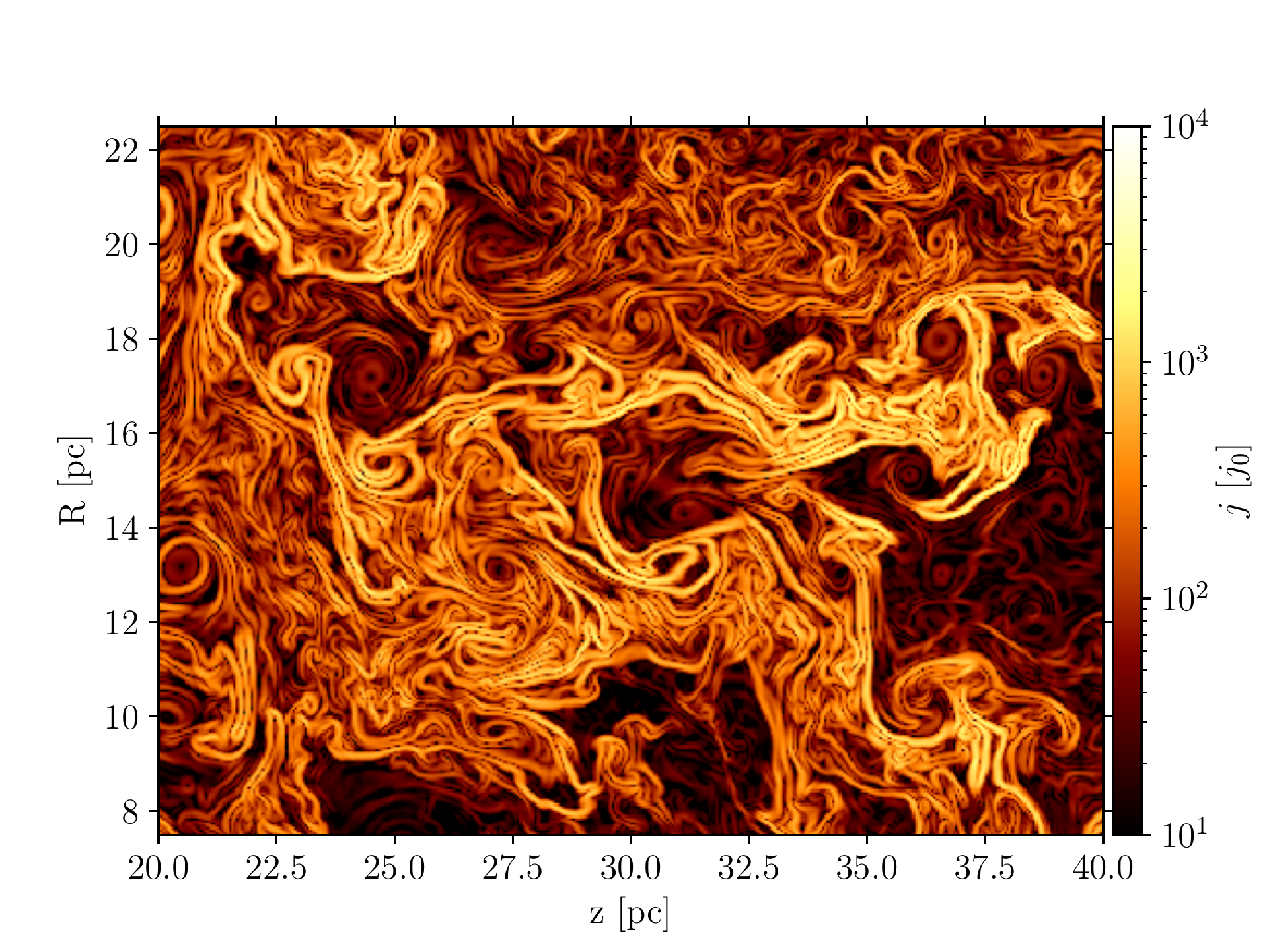}
  \end{minipage}
  \caption{ ({\it left}) Toroidal component of magnetic field ({\it top}) and magnetic energy ({\it Bottom}) maps of run L-R1 at $t = $ 88 kyrs. {(\it right)} Current density maps of run L-R1 at $t = $ 88 kyrs. The plot area corresponds to the green rectangle in the left panel. The two backflows, which are antiparallel toroidal fields, interact in the cocoon, which generates current sheets.}
   \label{fig:mag_field}
\end{figure*}

Figure \ref{fig:mag_field} ({\it left}) shows the distribution of toroidal magnetic fields and magnetic energy for run L-R1 at $t=88$ kyrs ({\it left}) and the distribution of current density, $j = \nabla \times B$ ({\it right}).
The magnetic energy is increased by shock compression; in particular, its energy is strongly enhanced downstream of the terminal shock of the western jet.
We find three areas in the cocoon in Figure \ref{fig:mag_field} ({\it left-top}):
an area of a positive field ({\rm blue}), an area of a negative field ({\rm red}), and an area of a weak magnetic field ({\rm white}).
Both backflows, which are antiparallel toroidal fields, interact in the cocoon, and the gas is mixed by vortices.
The magnetic fields are therefore dissipated by magnetic reconnection
in this mixing region.
Figure \ref{fig:mag_field} ({\it right}) shows that the interaction of both backflows generates current sheets, where particles can be efficiently accelerated by magnetic reconnection.
Our results would be affected by an artefact of the axisymmetric condition.
If we conduct three-dimensional simulations, turbulence would be generated in the cocoon, which would readily convert toroidal and poloidal fields.

\section{Discussion} \label{sec:discussion}
\subsection{Dynamics of Backflow} \label{sec:backflow}
We here make a simple estimation of the interval of the oblique shocks.
Note that the oblique shocks are formed by the interaction between the beam and backflow.
The interval thus corresponds to the size of the backflow vortex $l_{\rm bf}$.
For the eastern jet of L-R1, we assume that the radius $r_{\rm bf}$ is about 10 pc (see Figure \ref{fig:outline_tevo}).
To verify this assumption, we make the simple argument of mass flux conservation.
The simulation result gives a $z$-direction velocity and density of the backflow of about $0.15 v_{\rm jet}$ and $10^{-1}\rho_{\rm jet}$, respectively.
Conservation of mass flux in a hollow cylindrical symmetry gives $\pi R_{\rm jet}^2 \rho_{\rm jet} v_{\rm jet} = \pi( r^2_{\rm bf} - r^2_{\rm jet}) \rho_{\rm bf} v_{\rm bf}$.
We obtain $r_{\rm bf} \sim 8 R_{\rm jet} = 8$ pc, and the assumption is reasonable.
We assume that the $R$-direction velocity is about the speed of sound $c_{\rm bf}$.
To obtain $c_{\rm bf}$, we first calculate the speed of sound at the hotspot $c_{\rm h}$ using the Rankine--Hugoniot condition,
\begin{equation}
    c_{\rm h}^2 = \frac{5 \mathcal{M}^4 + 14\mathcal{M}^2 -3 }{16 \mathcal{M}^2} c_{\rm jet}^2,
\end{equation}
where $\mathcal{M}=10$ is the Mach number of the jets while $c_{\rm jet} = \mathcal{M}^{-1} v_{\rm jet}$ is the speed of sound of the jets.
Note that we ignore the advance speed because the advance speed is much lower than $v_{\rm jet}$ and $c_{\rm h}$.
We can estimate $c_{\rm bf}$ simply using the adiabatic condition:
\begin{equation}
    c_{\rm bf}^2 = \left( \frac{R_{\rm jet}^2}{r^2_{\rm bf}} \right)^{\gamma - 1} c_{\rm h}^2,
\end{equation}
where we assume that the radius of the hotspot is the same as the radius of the jet.
The $R$-direction velocity is therefore evaluated as
\begin{equation}
    c_{\rm bf} =  v_{\rm jet} \sqrt{ \left( \frac{R_{\rm jet}^2}{r^2_{\rm bf}} \right)^{\gamma - 1} \frac{5 \mathcal{M}^4 + 14\mathcal{M}^2 -3 }{16 \mathcal{M}^4}} \sim 0.12 v_{\rm jet}.
\end{equation}
This value is close to the $z$-direction velocity of backflow $v_{\rm bf}$, which is about $0.15 v_{\rm jet}$ (see Figure \ref{fig:l-r1_rovz}).
The oscillating time of the $R$-direction backflow is $\tau = r_{\rm bf}/ c_{\rm bf}$, and the interval of oblique shocks is thus evaluated as
\begin{equation}
    \label{eq:backflow}
     l_{\rm bf} = 2 \tau v_{\rm bf} \sim 25~{\rm pc}.
\end{equation}
This value is roughly consistent with our numerical result.
The bottom panel of Figure \ref{fig:shock} shows that the backflow departs from $z = 90$ pc and arrives at $z = 65$ pc, where a strong oblique shock is excited.
However, the actual dynamics of the beam and backflow are more complex and many weak shocks are thus excited along the beam.

\subsection{Synchrotron Emission}
\begin{figure*}
\begin{minipage}{0.5\hsize}
  \begin{center}
    \includegraphics[width=0.95\textwidth, bb = 0 0 460 345]{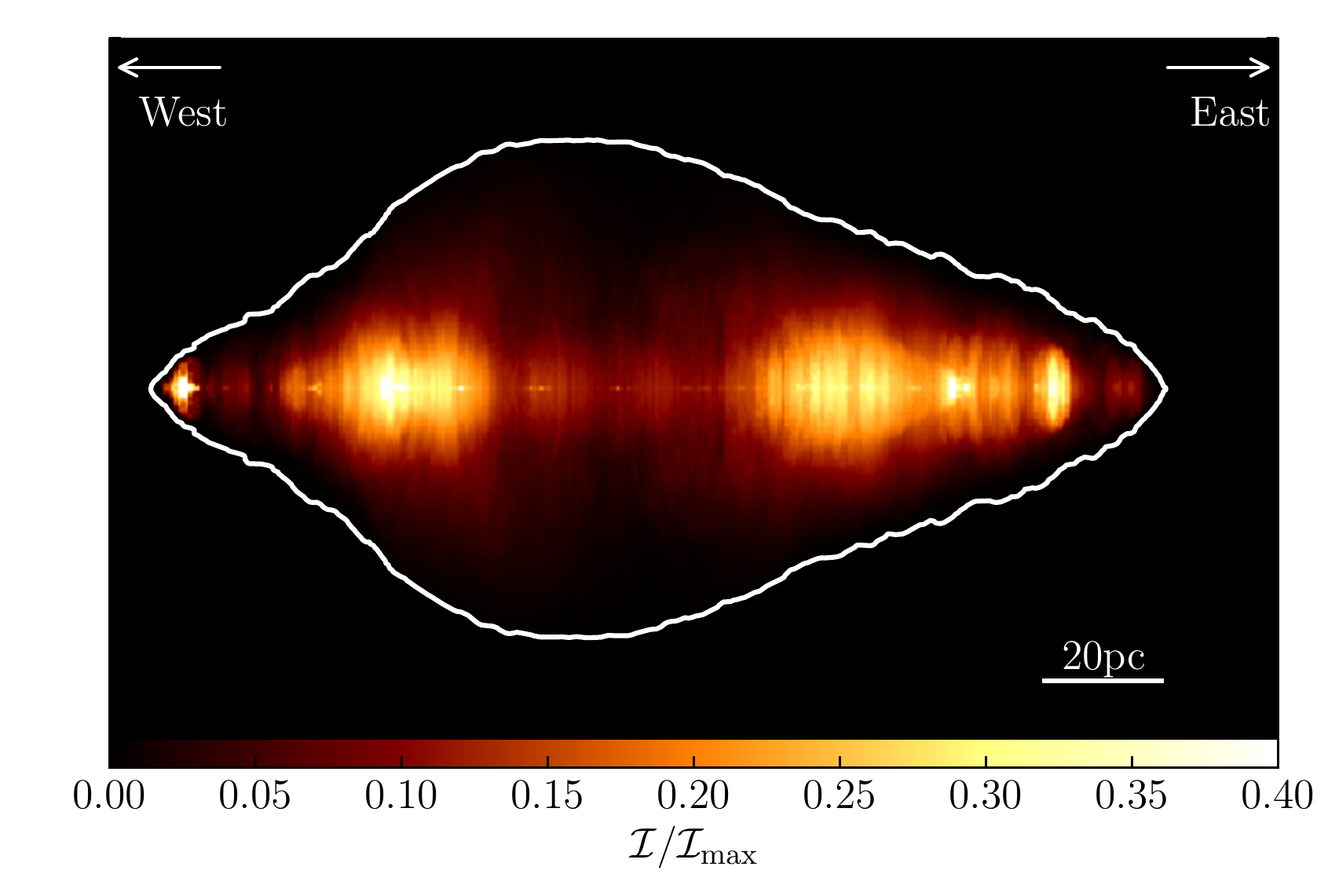}
  \end{center}
\end{minipage}
\begin{minipage}{0.5\hsize}
    \begin{center}
      \includegraphics[width=0.95\textwidth,bb=0 0 460 345]{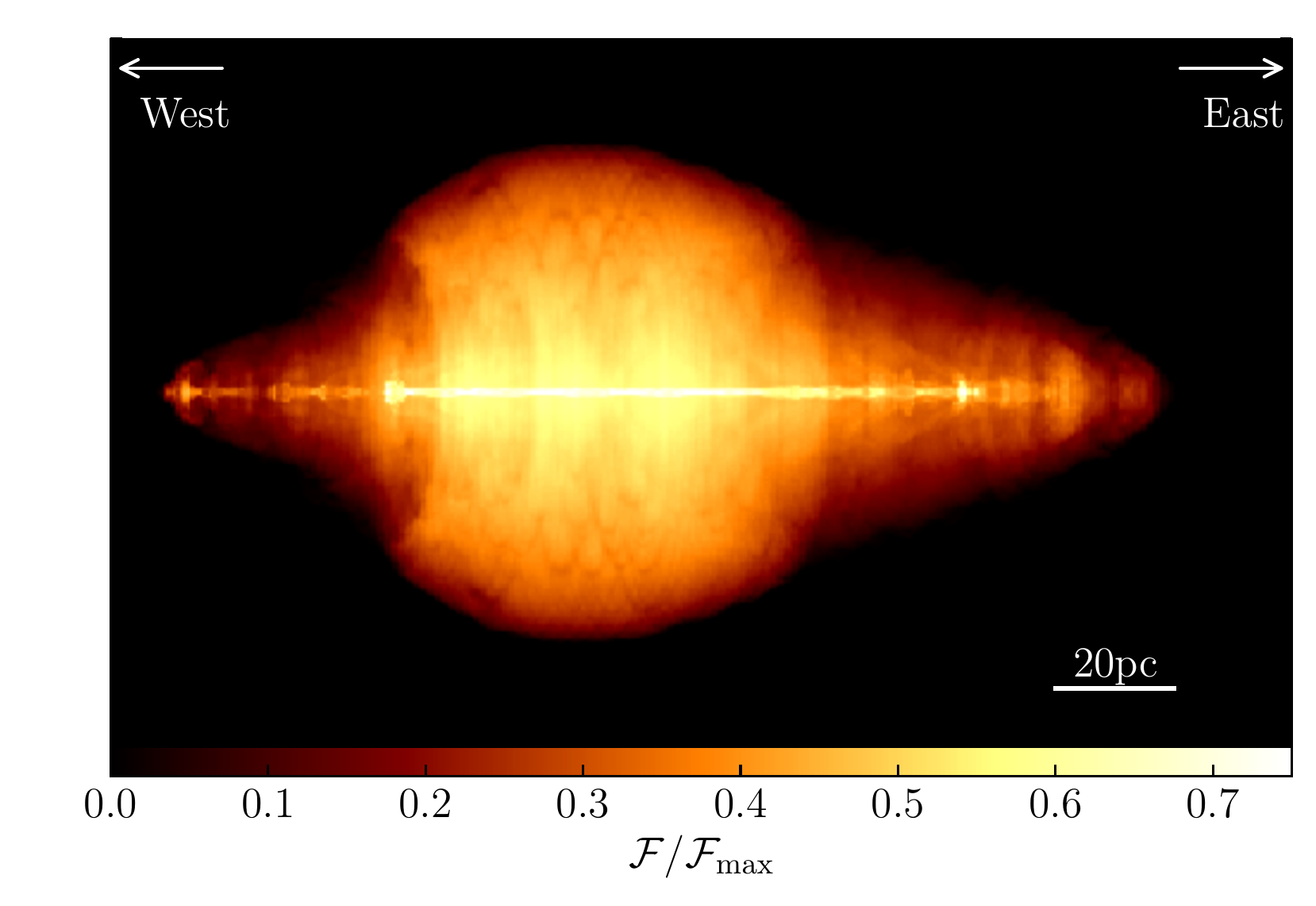}
    \end{center}
\end{minipage}
\caption{({\it left}) Synchrotron intensity map overlaid with the outer shape (white) of the jet in run L-R1 on a linear scale. ({\it right}) Map of the distribution of jet particles in run L-R1 on a linear scale. Both maps are plotted for a viewing angle of $78.8^\circ$}
\label{fig:Stokes_I}
\end{figure*}
We create a pseudo synchrotron image of W50/SS433 using physical values from MHD simulations.
To calculate the synchrotron emissivity, we convert the two-dimensional axisymmetric cylindrical coordinates to three-dimensional Cartesian coordinates and assume a viewing angle of $78.8^\circ$.
We calculate and integrate the emissivity at each cell along the line-of-sight direction:
\begin{equation}
    {\mathcal I} = \int_{LoS} {\mathcal J_{\rm sync}}ds,
\end{equation}
where ${\mathcal I}$ and ${\mathcal J_{\rm sync}}$ are respectively the synchrotron intensity and emissivity.
Our simulations do not model the evolution of non-thermal electrons, and hence some assumptions are needed to calculate the synchrotron emissivity.
One is that the spectral index is uniform everywhere.
The other is that the number density of non-thermal electrons is proportional to the gas density and thermal energy density.
Under these assumptions, the synchrotron emissivity is given by \citep{1996ApJ...472..245J}
\begin{equation}
  {\mathcal J_{\rm sync}} = C_1(\nu) \rho^{1-2\alpha} p^{2\alpha} B_{\perp}^{\alpha+1},
\end{equation}
where $C_1$, $B_{\perp}$, $\nu$, and $\alpha$ are a normalized constant, the magnetic field component in the plane of the sky, the frequency of radiation, and the spectrum index, respectively.
We adopt that $\alpha = 0.52$ because some radio observations provide the flatter spectrum at the eastern wing of W50 \citep{dubner1998,2011A&A...529A.159G}.
The calculation of polarized emissions is beyond the scope of this paper due to axisymmetry.

Figure \ref{fig:Stokes_I} ({\it left}) shows the synchrotron intensity map at $t = 88$~kyrs for run L-R1.
The synchrotron power strongly depends on magnetic energy, and a weak emission is thus observed along with the shell (see Figure \ref{fig:mag_field}).
The bright region in the western wing is around the oblique shocks and the termination shock.
Bright small spots on the jet axis are formed by oblique shock whose spacing is defined by equation \ref{eq:backflow}.

We can see the wide bright area both in the west and east wings which consist of thin filaments.
The accumulated magnetic field along the contact discontinuity is the origin of these bright regions.
We consider that these bright spots correspond to SS433 X-ray hot spots.
For example, the magnetic filament around (R, z) = (10, 80) is the origin of the bright filament closest to the eastern end.
On the other hand, because the magnetic field around the terminal shock is diffused, there is no X-ray radiation around that region.

The synchrotron intensity map obtained from numerical simulation is similar to X-ray observations of W50 \citep{1996A&A...312..306B,brinkmann1996,2007A&A...463..611B,1997ApJ...483..868S}.
In particular, the positions of X-ray hotspots of W50 (e1, e2, w1, and w2) are consistent with those in the numerical results.
The oblique shocks are thus thought to be rather efficient particle accelerators in contrast with the terminal shocks in the long-term evolution of jets.
\cite{2020ApJ...889..146S} and \cite{2020ApJ...904..188K} constructed an analytic model for non-thermal leptonic and hadronic emission from knots in SS433 jets, and their results suggest that the X-ray observation can be reproduced by a synchrotron emission when the acceleration process is efficient.

We next discuss the distribution of electrons emitting radio waves in W50/SS433.
One main difference between microquasar jets and AGN jets is the outburst age; i.e., the age of relativistic electrons in the jet cocoon.
The synchrotron cooling time $t_{\rm c}$ of non-thermal electrons in the microquasar jets is $t_{\rm c} \sim 1.9 \times 10^{13} B^{-2} E_{\rm e}^{-1}~{\rm s}$, where $E_{\rm e}$ is the energy of electrons.
The magnetic field strength of the eastern and western wings is about $1--100~{\rm \mu G}$ \citep{1997ApJ...483..868S} and the cooling time scale is thus estimated as $t_{\rm c} \sim 100~{\rm kyrs}$ when the frequency is $1~{\rm GHz}$ and the magnetic field is $100~{\rm \mu G}$.
The plasma created around the terminal or oblique shock is thus not cooled to a few GeV and emits radiation at a centimeter wavelength.
We calculate the shocked particle content in W50/SS433, ${\mathcal F}$, to integrate the tracer function, $f$:
\begin{equation}
  {\mathcal F} = \int_{LoS} f ds.
\end{equation}
Figure \ref{fig:Stokes_I} ({\it right}) is the same as Figure \ref{fig:Stokes_I} ({\it left}) but we display a map of the distribution of jet particles .
The plasma accumulates in the cocoon after passing the terminal shock and oblique shocks.
The lifetime of electrons for synchrotron radiation is much longer than the jet active time (simulation time), and relativistic electrons in the shell can thus emit continually.
The intensity is lower in both wings than in the shell because relativistic electrons are advected by backflow so that there are fewer electrons in both wings.
This characteristic is also found for the radio emission of the eastern wing.
Another radio characteristic of the eastern wing is an extended radio filament in the terminal region.
In addition, \cite{2007A&A...463..611B} found an X-ray ring in this region, which corresponds spatially to the terminal shock of the SS433 jet.
However, our numerical result indicates that the shock size is only a few parsecs.
The filament is thus formed by the backflow of the terminal shock.
Additionally, \cite{2018PASJ...70...27S} reported the existence of helical fields that coil the eastern wings.
This result is consistent with our backflow scenario.
We next discuss the shell emissions of W50.
We find that magnetic reconnection generates current sheets in the shell (see Figure \ref{fig:mag_field}) and re-accelerated electrons in current sheets provide a prominent radio emission.
This radio emission corresponds to the filament structure in the southern region of W50; otherwise, the MeV $\gamma$-ray emission observed in the northern region of W50 is explained by this acceleration scenario.
In this simulation, we assume an axisymmetric condition and the magnetic field in the shell of W50 is thus exhausted by magnetic reconnection.
However, if we carry out three-dimensional simulations, the magnetic field can be stored in the shell because the flow can escape easily in the $4\pi$ direction.
The synchrotron emission from the shell would therefore be enhanced.

\subsection{SNR Evolution}　\label{sec:discussion_snr}
\begin{figure}
    \includegraphics[width=1.0\columnwidth,bb=0 0 461 346]{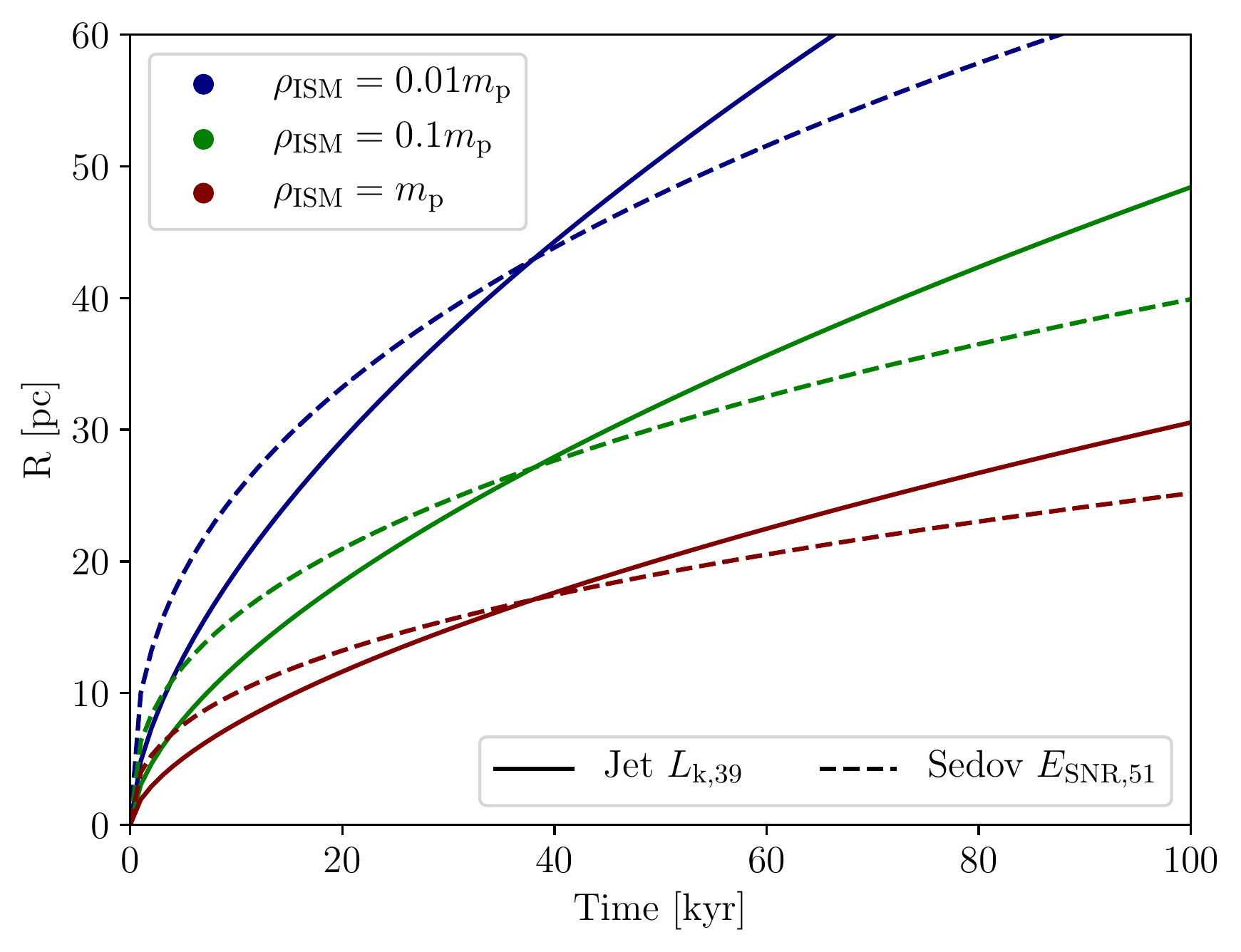}
    \caption{Radial distance for SNRs ({\it dashed}) and jets ({\it solid}) as a function of time in the case of ISM gas densities $\rho_{\rm ISM} = 0.01~m_{\rm p}~({\rm blue}),~0.1~m_{\rm p}~({\rm green}),$and$ ~1~m_{\rm p}~({\rm red})~{\rm g~cm^{-3}}$.
    We assume that the kinetic energy luminosity of jets and the initial explosion energy of SNRs are $L_{\rm k,39} = 10^{39}$ erg ${\rm s^{-1}}$ and $E_{\rm SNR,51} = 10^{51}$ erg, respectively.
    }
    \label{fig:snr_vs_jet_radial_evo}
\end{figure}
Previous studies have suggested that the SNR produced by the compact object SS433 formed the shell structure of W50, and we thus discuss the difference in radial expansion between our backflow scenario and SNR scenarios.
We next compare the features of radio emissions of W50/SS433 with those of other SNRs.
We mention again the shell diameter is $D = 96~{\rm pc}$.
The typical evolution of an SNR is described in the Sedov phase (i.e., the adiabatic phase).
An expanding SNR is then prescribed by the snowplow phase when radiative losses become important.
We assume that the shell of W50 is in the Sedov phase and radiative cooling is thus negligible.
Note that the radial expansion of the Sedov phase is faster than that of the snowplow phase.
The shell radius of the SNR develops in the Sedov phase as $R \propto (E_{\rm SNR}t^2/\rho_{\rm ISM})^{1/5}$, where $E_{\rm SNR}$ is the initial explosion energy of the SNR.
The dashed lines in Figure \ref{fig:snr_vs_jet_radial_evo} show the radius of the SNR as a function of time for three different values of the ISM density.
We simply assume that the ISM has a constant density  $\rho_{\rm ISM} = 1~m_{\rm p}, 0.1~m_{\rm p},$ and $0.01~m_{\rm p}~{\rm g~cm^{-3}}$ and $E_{\rm SNR, 51} = 10^{51}~{\rm erg}$.
When the ISM density is greater than $0.1 m_{\rm p}$, it takes more than 100 kyrs to reach a shell radius of 48 pc.
The ISM is thus about $0.01 m_{\rm p}$, which is a very low density with which to form the W50/SS433 shell.
However, we note that W50/SS433 is located near the galactic plane, and HI emissions are observed around W50 \citep{2021arXiv210305578S,2011ApJS..194...20P}.
Such a low-density environment is thus unlikely.
In another case, we believe that the W50 nebula formed in a high-power explosion $(E_{\rm SNR} \gg 10^{51}~{\rm erg})$, such as a hypernova or multiple supernovas.
In contrast, we plot the time evolution of the radial expansion for the light jets for jet kinetic energy luminosity $L_{\rm kin, 39} = 10^{39}~{\rm erg~s^{-1}}$ in Figure \ref{fig:snr_vs_jet_radial_evo}.
The approximate solution of the radial expansion of jets is $R_{\rm jet} \propto t^{3/5}$.
We mention again the Sedov solution is proportional to $t^{2/5}$.
The radius obtained using the jet model becomes larger than that obtained using the SNR model when the total energy exceeds the SN explosion energy, $E_{\rm SNR}=10^{51}~{\rm erg}$.
The time is about $t \sim 40$~kyrs for these model settings.
In the case of jets instead of an SNR, the shell of W50 reaches $48$~pc within 100~kyrs when the ISM is $0.1 m_{\rm p}~{\rm g~cm^{-3}}$.
Therefore, jets having long-term activity, for which a lifetime of $> 10$~kyrs is expected, are a good candidate for the formation mechanism of W50/SS433.

\begin{figure}
    \includegraphics[width=1.0\columnwidth,bb=0 0 461 346]{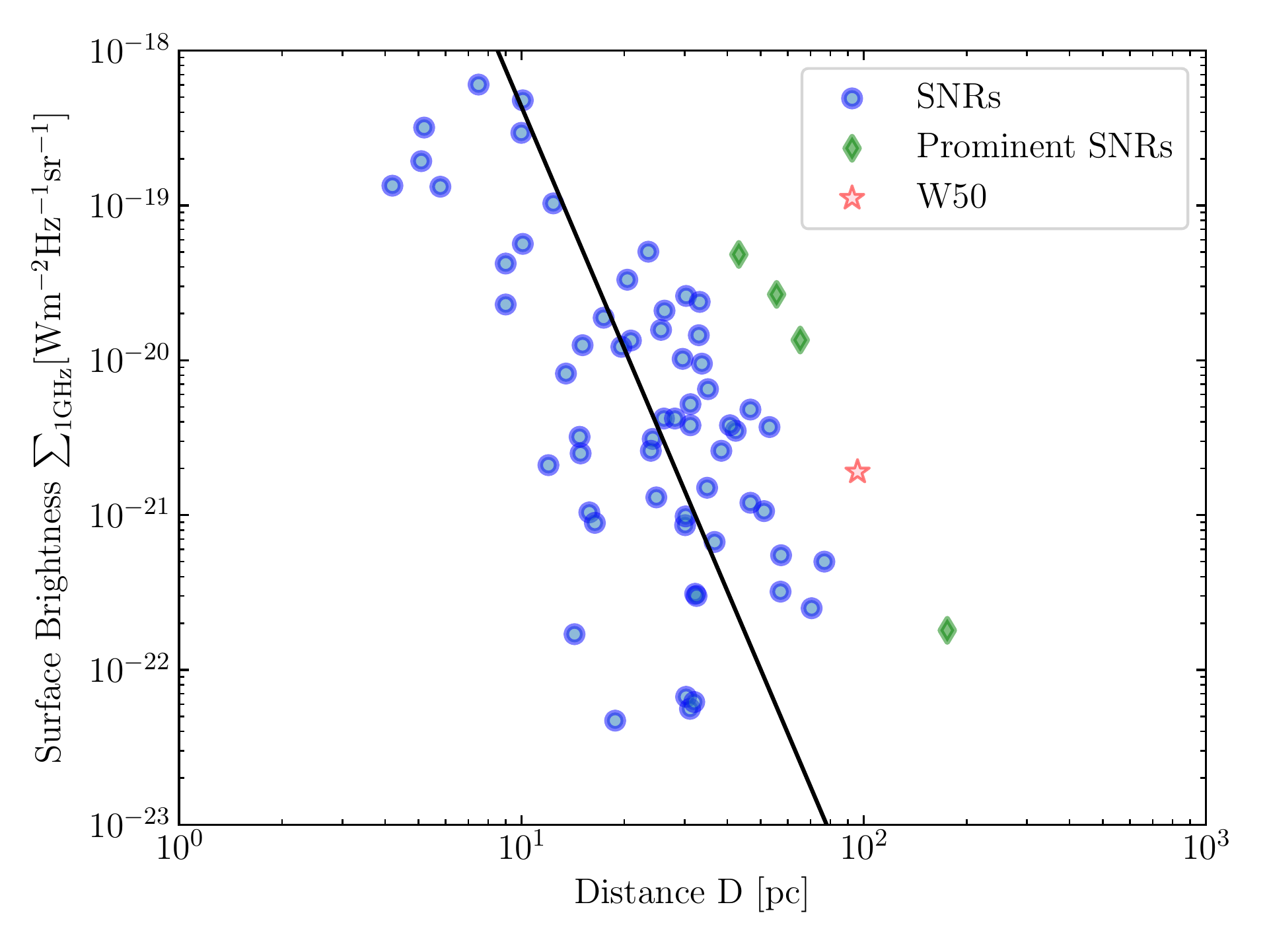}
    \caption{Surface brightness vs. diameter at 1 GHz for shell SNRs (blue circles and green diamonds) in \cite{2014SerAJ.189...25P} and W50 (red star). The solid line is the orthogonal regression in \cite{2014SerAJ.189...25P}. Green diamonds are four prominent galactic SNRs (CTB 37A, Kes97, CTB 37B, and G65.1+0.6), which interact with molecular clouds.}
    \label{fig:snr_d-sig}
\end{figure}

We next discuss radio observations of typical SNRs and W50/SS433.
Galactic SNRs have a radio observational relationship between the radio surface brightness and the diameter of the SNR, the so-called radio surface-brightness-to-diameter relationship ($\Sigma_{\rm R}-D$) \citep[e.g.,][]{2014SerAJ.189...25P}.
Figure \ref{fig:snr_d-sig} displays the $\Sigma_{\rm R}-D$ relationship at $1~{\rm GHz}$ for galactic SNRs (blue circles and green diamonds) and W50/SS433 (red star).
The figure shows that W50/SS433 is prominent and has a large size compared with galactic SNRs.
We note the observations of four prominent galactic SNRs (CTB 37A, Kes97, CTB 37B, and G65.1+0.6) interacting with molecular clouds, which enhances radio emissions.
The wings of W50 have a radio surface brightness comparable to that of the shell of W50 \citep{dubner1998}.
Previous hydrodynamical simulations clearly showed that the jetted gas and the gas-related supernova explosions are separated; in particular, the jetted gas is not distributed around the shell region \citep[e.g.,][]{2011MNRAS.414.2838G}.
Although some observations suggest that there are molecular clouds associated with jets in the western wing of W50 \citep{2008PASJ...60..715Y,2020ApJ...892..143L}, there is no critical evidence of interacting molecular clouds, which would accelerate non-thermal particles.
Additionally, the jet–cloud interaction is not strong and it thus does not enhance radio emissions throughout the western wing.
Therefore, the origin of the prominent emission from the shell of W50 is not the same as that of other prominent SNRs.

Finally, we mention that the shell of W50 maintains a nearly perfectly circle-like structure.
\cite{2019ApJ...884..113S} investigated the relationship between the radio morphology and the sizes (ages) of SNRs.
They found that larger (older) SNRs are more elliptical and/or elongated because the inhomogeneous ISM, turbulent structure, and molecular clouds affect their morphology.
As we mentioned above, W50 is in a large size (old) category if the shell of W50 comprises SNRs.

\subsection{Comparison with other W50/SS433 formation scenarios} \label{sec:4.3}
In this section, we compare our results with those of \cite{2011MNRAS.414.2838G}.
We first mention that the radial expansion of the SNR is too fast in the simulations conducted by \cite{2011MNRAS.414.2838G} compared with the Sedov solutions and hydrodynamical simulations of SNRs.
They assumed the kinetic energy of SNRs is $\sim 10^{51}~{\rm erg}$ and the ISM density is $\sim 0.1~{\rm cm^{-3}}$, and the shell radius of SNRs reaches $50~{\rm pc}$ within $20~{\rm kyrs}$.
They might have underestimated the normalization of the kinetic energy of SNRs or overestimated the density unit of the ISM density by an order of magnitude.
We also mention the advance speed of jets in \cite{2011MNRAS.414.2838G}.
As an example, model Jet1 in \cite{2011MNRAS.414.2838G} has a young age of 2.2~$\rm kyrs$ when the jet crosses the tip of the eastern wing.
The average advance speed of model Jet1 almost maintains the launch velocity of $0.26c$ from SS433; i.e., $120~{\rm pc}/2.2 ~{\rm kyrs} \sim 0.18c$.
Thus, the advance of model Jet1 does not decelerate.
Note that the results of kinematical studies of the radio observations of W50 indicate that an upper limit to the proper motion of the radio filaments of the eastern wing is $0.0405c$ \citep{2011MNRAS.414.2828G}.
Jets that do not decelerate must be comparable or greater than the ISM in density.
Some previous numerical simulations have modeled protostar jets for heavy jets \citep[e.g.,][]{2016A&A...596A..99U}.
However, numerical and observational studies of AGN jets and microquasar jets suggest that extensive cocoons develop when the surrounding medium is lighter than the medium of the jets.

\cite{2017A&A...599A..77P} constructed a model combined with the theory of the dynamics of SNRs, wind bubbles, and jets for the formation of W50.
Their jet model was based on turbulent jets in \cite{1987flme.book.....L}.
In the case of a wind bubble, if the binary system of SS433 is a super-Eddington accretion, the age of the wind bubble needs to exceed $300~{\rm kyrs}$ to produce the large shell of W50.
They also modeled the dynamics of SNRs in the Sedov phase and snowplow phase, describing the dynamics using the fitting function of \cite{2015ApJ...802...99K}.
They argued that the age of W50 as an SNR is at least $97~{\rm kyrs}$, which agrees with our results in Figure \ref{fig:snr_vs_jet_radial_evo}.
The main result of \cite{2017A&A...599A..77P} is that the shell of W50 formed in a supernova explosion about 100 kyrs ago, and the SS433 jets produce both wings and pushed up the pre-existing SNR within $27~{\rm kyrs}$.
However, hydrodynamical studies showed that interaction between the jets and SNR does not push up the shell of W50 in the radial direction \citep{2011MNRAS.414.2838G}
Therefore, some other mechanism is needed for the model of \cite{2017A&A...599A..77P} to exchange energy efficiently between the jets and SNR shell.
\section{Summary} \label{sec:summary}
We conducted a series of axisymmetric MHD simulations of two-side jets to study the parameter dependence for the formation of shell and wings of the W50/SS433 system.
We argue that light and supersonic jets can produce the shell and wing morphology in a backflow scenario.
The main results of this work are as follows.
\begin{itemize}
  \item  At an early time ($t \lesssim 30~{\rm kyrs}$), the morphology of light jets ($ \eta < 10^{-3} $) had a spheroidal shape. Afterward, the shell and wings formed and they expanded under self-similar evolution.
  In contrast, slightly heavy jets ($\eta \sim 10^{-2}$) cannot form a large shell because the advance is too fast compared with the radial expansion.
  \item Jets having a large radius do not decelerate like jets having a smaller radius. Moreover, larger jets provide much kinetic energy luminosity, and the radial expansion of the cocoon thus becomes fast. We can therefore produce wings and a shell of arbitrary size by appropriately selecting parameters, such as the jet radius and density contrast.
  \item  The exponential ISM profile produces the asymmetric wings observed for W50/SS433 in previous works \citep{2008MNRAS.387..839Z, 2011MNRAS.414.2838G, 2019ASSP...55...71M}.
  We find that the length ratio between the eastern and western wings, $l_{\rm east}/l_{\rm west}$ are roughly the same for all runs when the positions of jets are fixed. Thus, $l_{\rm east}/l_{\rm west}$ determined only by the density profile of ISM.
  Furthermore, our runs using the ISM profile of our model provide an asymmetric shape that is consistent with observations.

  \item The main challenge to the jet scenario is the requirement for the long-period activity of the jets. Run L-R1 requires a long activity lasting more than $100~{\rm kyrs}$ to result in the observed size of W50.
  The advance speed and the speed of radial expansion respectively depend on the density contrast and density of the ISM.
  Therefore, if the ISM is slightly lower than $0.1~{\rm cm^{-3}}$, our scenario overcomes the issues noted above.
  \item A large amount of jet kinetic energy is converted into thermal energy for jets propagating in the galactic plane because these jets interact with the high-density ISM.
  The interaction between the backflows and beams drive strong oblique shocks at the intermediate point of beams. The positions of oblique shocks correspond to X-ray hotspots of W50. Meanwhile, highly turbulent flows drive magnetic reconnection, which generates current sheets in the shell region where the two backflows have interacted.

  \item We discussed the jet scenario and SNR scenario.
    For an SNR interpretation, an extremely-low-density atmosphere, $n_{\rm ISM} \ll 0.1~{\rm cm^{-3}}$, or a high-energy explosion, $E_{\rm SNR} \gg 10^{51}~{\rm erg}$, is required for the SNR's froward shock to reach about $50~{\rm pc}$. Moreover, the shell of W50 is larger and much more prominent than that of other galactic SNRs.
    Although the jet scenario strongly depends on the activity time and the total amount of energy, the shell of W50 could have been formed by backflow within a period of 100 kyrs. Notice that our numerical results can explain the morphology of W50/SS433, but they do not entirely rule out the SNR scenario.
\end{itemize}
Future work should treat some of the key physics that we ignored.
First, three-dimensional effects affect the dynamics of jets.
The advance speeds in three-dimensional simulations are higher than those in axisymmetric simulations, and turbulence is generated by non-axisymmetric motions.
Second, we ignored winds from the companion star and/or accretion disk.
The interaction between winds and jets might positively affect the formation of the shell of W50.
If SS433 is an ultraluminous X-ray source, then the state of the accretion disk is a supercritical accreting flow.
Radiation hydrodynamics simulations of supercritical accretion flow suggest that strong winds ($\sim 10^{38}~{\rm erg~s^{-1}}$) can be driven by radiation pressure \citep{2009PASJ...61..769K}.
From an observational viewpoint, some ultraluminous X-ray sources have bubbles on the scale of a few tens of parsecs, similar to the case for W50 \citep[e.g.,][]{2010Natur.466..209P, 2020ApJ...896..117B}.
Finally, we need to implement the precession of jets.
However, we mention again that the precessions do not affect the large-scale morphology of W50 because the jets are well collimated within 0.1 pc from SS433 and they become hollow \citep{2019ASSP...55...71M}.
\acknowledgments We thank the anonymous referee for comments that have improved the clarity of this paper. We are grateful to Dr. H., Yamamoto for useful discussion.
This work was supported by JSPS KAKENHI Grant Numbers JP20J12591 (T.O.), JP19K03916, JP20H01941 (M.M.) and JP20J13339 (H.S.).
Our numerical computations were carried out on the Cray XC50 at the Center for Computational Astrophysics of the National Astronomical Observatory of Japan.
The computation was carried out using the computer resource by Research Institute for
Information Technology, Kyushu University.
We thank Glenn Pennycook, MSc, from Edanz Group (https://en-author-services.edanzgroup.com/ac) for editing a draft of this manuscript.
\bibliographystyle{aasjournal}
\bibliography{ref}

\end{document}